\documentclass[onecolumn]{aastex62}
\usepackage{amsmath}     
\usepackage{wasysym}           
\usepackage{graphicx}
\usepackage{amssymb}
\usepackage{epstopdf}
\usepackage{mathrsfs}
\usepackage{anyfontsize}
\usepackage{natbib}
\usepackage{color}
\usepackage{lipsum}
\usepackage{diagbox}

\shorttitle{Gravitational Instability of Adiabatic Filaments}
\shortauthors{Coughlin \& Nixon}
\begin{document}
\title{The Gravitational Instability of Adiabatic Filaments}
\author[0000-0003-3765-6401]{Eric R.~Coughlin}
\affiliation{Department of Astrophysical Sciences, Princeton University, Princeton, NJ 08544, USA}
\author[0000-0002-2137-4146]{C.~J.~Nixon}
\affiliation{Department of Physics and Astronomy, University of Leicester, Leicester, LE1 7RH, UK}

\email{eric.r.coughlin@gmail.com}

\begin{abstract}
Filamentary structures, or long and narrow streams of material, arise in many areas of astronomy. Here we investigate the stability of such filaments by performing an eigenmode analysis of adiabatic and polytropic fluid cylinders, which are the cylindrical analog of spherical polytropes. We show that these cylinders are gravitationally unstable to perturbations along the axis of the cylinder below a critical wavenumber $k_{\rm crit} \simeq few$, where $k_{\rm crit}$ is measured relative to the radius of the cylinder. Below this critical wavenumber perturbations grow as $\propto e^{\sigma_{\rm u}\tau}$, where $\tau$ is time relative to the sound crossing time across the diameter of the cylinder, and we derive the growth rate $\sigma_{\rm u}$ as a function of wavenumber. We find that there is a maximum growth rate $\sigma_{\rm max} \sim 1$ that occurs at a specific wavenumber $k_{\rm max} \sim 1$, and we derive the growth rate $\sigma_{\rm max}$ and the wavenumbers $k_{\rm max}$ and $k_{\rm crit}$ for a range of adiabatic indices. To the extent that filamentary structures can be approximated as adiabatic and fluid-like, our results imply that these filaments are unstable without the need to appeal to magnetic fields or external media. Further, the objects that condense out of the instability of such filaments are separated by a preferred length scale, form over a preferred timescale, and possess a preferred mass scale.
\end{abstract}

\section{Introduction}
The formation and evolution of filamentary structures is ubiquitous in astrophysics. To give a handful of examples, the first large scale structures are thought to form out of the intersections and collapse of dark matter sheets and filaments (e.g., \citealt{bond96, kravtsov12}). Galaxy mergers and the tidal destruction of dwarf galaxies are accompanied by the formation of galactic tidal tails and stellar streams (e.g., \citealt{grillmair09}). The tidal disruption of a star by a supermassive black hole \citep{rees88} transforms the star into a long, thin filament of gas \citep{kochanek94, guillochon14, coughlin16}. The merger of two compact objects results in the tidal stripping of material from the less dense object, and that material is flung out in the form of a tidally ejected tail of debris (e.g., \citealt{lee07, rosswog07}). Mergers between giant molecular clouds and cloud cores, and the associated turbulence within those cores, results in the formation of filaments along which star formation can occur (e.g., \citealt{andre10, andre17}). Cold streams can provide a reservoir of gas that fuels star formation in galaxies (e.g., \citealt{dekel06}). 

Owing to their ubiquity, the (magneto)hydrodynamic stability of such filaments has received a considerable amount of attention, with among the first analytic investigations undertaken by \citet{chandrasekhar53} and \citet{ostriker64a} who analyzed the stability of an incompressible and uniform-density cylinder and a compressible and uniform-density cylinder, respectively. Since then, a number of authors have considered the stability of filaments in various environments with a variety of initial conditions and perturbations, which can give rise to, e.g., the Kelvin-Helmholtz instability (e.g., \citealt{mandelker16, padnos18, vossberg19, mandelker19}) and cooling instabilities (e.g., \citealt{bodo93, bessho12, birnboim16, gronke19, mandelker20}), which themselves can be modified by magnetic fields (e.g., \citealt{nagasawa87, berlok19}), rotation (e.g., \citealt{freundlich14, sadhukhan16}), and gravity (e.g., \citealt{nagasawa87, hunter97, hunter98, aung19}). 

In spite of the past work on the stability of filamentary structures, to the authors' knowledge only \citet{breysse14} analyzed the global stability of an adiabatic, hydrostatic cylinder with a polytropic equation of state through an eigenmode analysis (analogous to a stellar eigenmode analysis; e.g., \citealt{cox80}). \citet{breysse14} demonstrated that such filaments are unstable to (cylindrical) radial perturbations once the adiabatic index falls below one, in agreement with \citet{chandrasekhar53}. They also showed that the $g$-modes of such cylinders are convectively unstable to non-radial perturbations when the adiabatic index is less than the polytropic index, similar to the stellar convective instability criterion. \citet{breysse14} also focused on perturbations that displace the axis of the cylinder ($m = 1$ modes; see Section \ref{sec:general} for a precise definition of $m$), which are analogous to spherical perturbations that are characterized by a spherical harmonic number of $\ell = 1$ that displace the center of mass of the sphere.

Here we perform an eigenmode analysis of adiabatic fluid cylinders. We construct and analyze a purely Eulerian set of perturbation equations from the fluid equations and consider the fluid velocity as the fundamental variable, which contrasts the approach taken in most studies of stellar oscillations in which a Lagrangian formalism is adopted that treats the fluid displacement as the fundamental variable. We focus exclusively on purely polytropic cylinders, for which the adiabatic and polytropic indices are equal and g-modes are absent, and we primarily consider perturbations that are only along the axis of the cylinder and are azimuthally symmetric (i.e., no variation around the axis of the cylinder). 

In Section \ref{sec:hse} we derive the equilibrium configuration of a polytropic, hydrostatic cylinder, and in Section \ref{sec:cylperts} we analyze the response of such a cylinder to (cylindrical-) radial perturbations. We derive the same result as \citet{chandrasekhar53} and \citet{breysse14}, that such perturbations drive an instability when the adiabatic index of the gas satisfies $\gamma \le 1$; we emphasize, however, that as the gas becomes more isothermal and $\gamma \rightarrow 1$, this instability is fundamentally different from the gravitational instability of a $\gamma \rightarrow 4/3$ spherical polytrope. We present the generic (i.e., not restricted to the radial direction) perturbation equations and we derive the eigenvalue equations that describe the fundamental modes of the filament\footnote{The majority of the (abridged) algebraic manipulations that lead to the equations are given in Appendices \ref{sec:cylderivation} and \ref{sec:derivation}} in Section \ref{sec:general} .

In Section \ref{sec:m0modes} we consider perturbations that are azimuthally symmetric and along the axis of the cylinder, and we show that such perturbations are characterized by an unstable mode when the wavelength of the perturbation is longer than a critical value. We derive the growth rate of the instability as a function of the wavenumber of the perturbation for a number of different adiabatic indices, and we show that there is a maximum growth rate at a wavelength that is comparable to the radius of the cylinder. We show that this unstable mode characterizes unidirectional motion along the axis of the cylinder, and that the instability operates gravitationally and analogously to that of a $\gamma = 4/3$, spherical polytrope. We also present an eigenmode decomposition of a specific, initial perturbation. 

We summarize and discuss the implications of our findings in Section \ref{sec:summary}.

\section{Hydrostatic, polytropic cylinders}
\label{sec:hse}
Our unperturbed, hydrostatic state is a cylinder of length $2L$ and cross-sectional radius $H$ such that $L \gg H$, and we assume that fluid quantities (e.g., the density and pressure) only depend on cylindrical radius $s$, which is consistent with the hydrostatic nature of the cylinder in the limit that $L \rightarrow \infty$. Defining the line mass of the cylinder\footnote{This definition of the line mass is algebraically convenient, but is somewhat of a misnomer owing to the fact that the total length of the cylinder is $2L$; $\Lambda$ as we have defined it is therefore twice of the total line mass of the cylinder, being the total mass divided by the total length. } as $\Lambda = M/L$ and the dimensionless cylindrical radius as $\xi = s/H$, we parameterize the density $\rho$, the dimensionless line mass $\lambda$, and the gas pressure $p$ by

\begin{equation}
\rho = \frac{\Lambda}{4\pi H^2}g_0(\xi), \quad \lambda_0(\xi) = \int_0^{\xi}g_0(\tilde{\xi})\tilde{\xi}\,d\tilde{\xi}, \quad p = \frac{G\Lambda^2}{4\pi H^2}h_0(\xi), \label{rhodef}
\end{equation}
from which it follows that $\lambda_0(1) = 1$; here subscript zeros refer to the unperturbed state, which is relevant for the next two sections in which we consider perturbations on top of this state.

We further assume that the cylinder is a polytrope, implying that the functions $h_0$ and $g_0$ are related by $h_0 = K_0 g_0^{\gamma}$, where $K_0$ is the dimensionless specific entropy of the cylinder and $\gamma$ is the polytropic index. With these definitions, the equation of hydrostatic balance in the radial direction and the Poisson equation in cylindrical coordinates (see Equation \ref{contappA}) can be combined to yield the following, single equation for $\lambda_0$

\begin{equation}
\frac{K_0\gamma}{\gamma-1}\frac{\partial}{\partial \xi}\left[\left(\frac{1}{\xi}\frac{\partial \lambda_0}{\partial \xi}\right)^{\gamma-1}\right] = -\frac{\lambda_0}{\xi}, \label{hse}
\end{equation}
which is just the cylindrical version of the Lane-Emden equation written it in terms of the line mass $\lambda_0$. It is straightforward to show that if the solution to this equation is to be non-trivial and satisfy the requirement that the total line mass be equal to $\Lambda/2$, so that $\lambda_0(1) = 1$, then the function $\lambda_0$ can be approximated near the surface by

\begin{equation}
\lambda_0 \simeq 1-\frac{\gamma-1}{\gamma}\left(\frac{\gamma-1}{K_0\gamma}\right)^{\frac{1}{\gamma-1}}\left(1-\xi\right)^{\frac{\gamma}{\gamma-1}}\left\{1-\frac{\gamma}{2\gamma-1}\frac{\gamma-3/2}{\gamma -1}\left(1-\xi\right)\right\}. \label{msurf}
\end{equation}
To arrive at this expression, we Taylor expanded the function $\lambda_0$ in powers of $(1-\xi)$ and equated like powers on the left and right-hand sides of Equation \eqref{hse}. We can use this solution to integrate Equation \eqref{hse} inward from a point near the surface. We require that $\lambda_0$ go to zero at $\xi = 0$, which fixes the value of $K_0$ for a given $\gamma$. We determine $K_0$ numerically through an iterative method, specifically by integrating Equation \eqref{msurf} from $\xi \simeq 1$ to $\xi \simeq 0$ and varying $K_0$ until we achieve $\lambda_0(\xi \simeq0) \simeq 0$.

\begin{figure}[htbp] 
   \centering
   \includegraphics[width=0.495\textwidth]{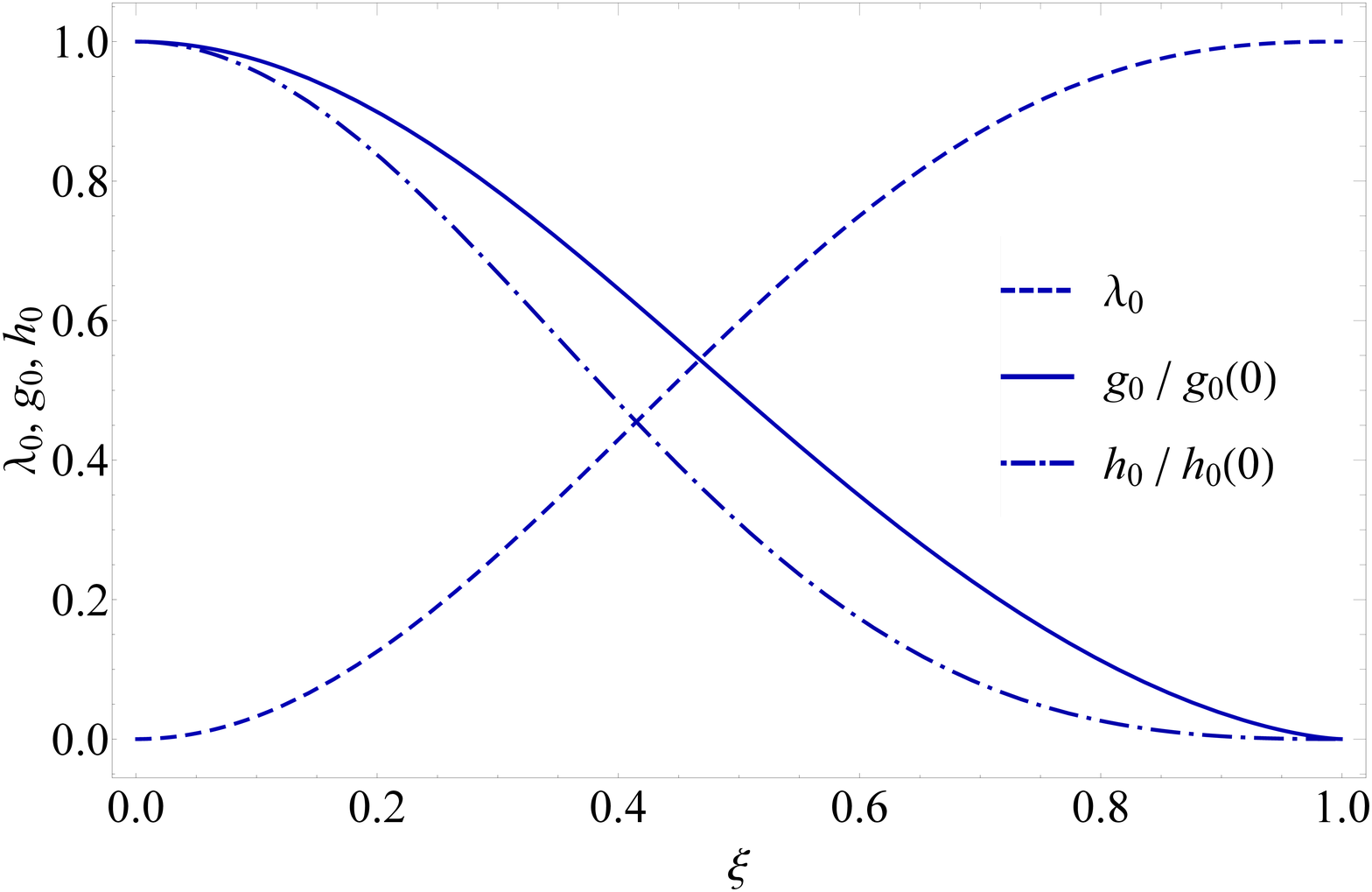} 
   \includegraphics[width=0.495\textwidth]{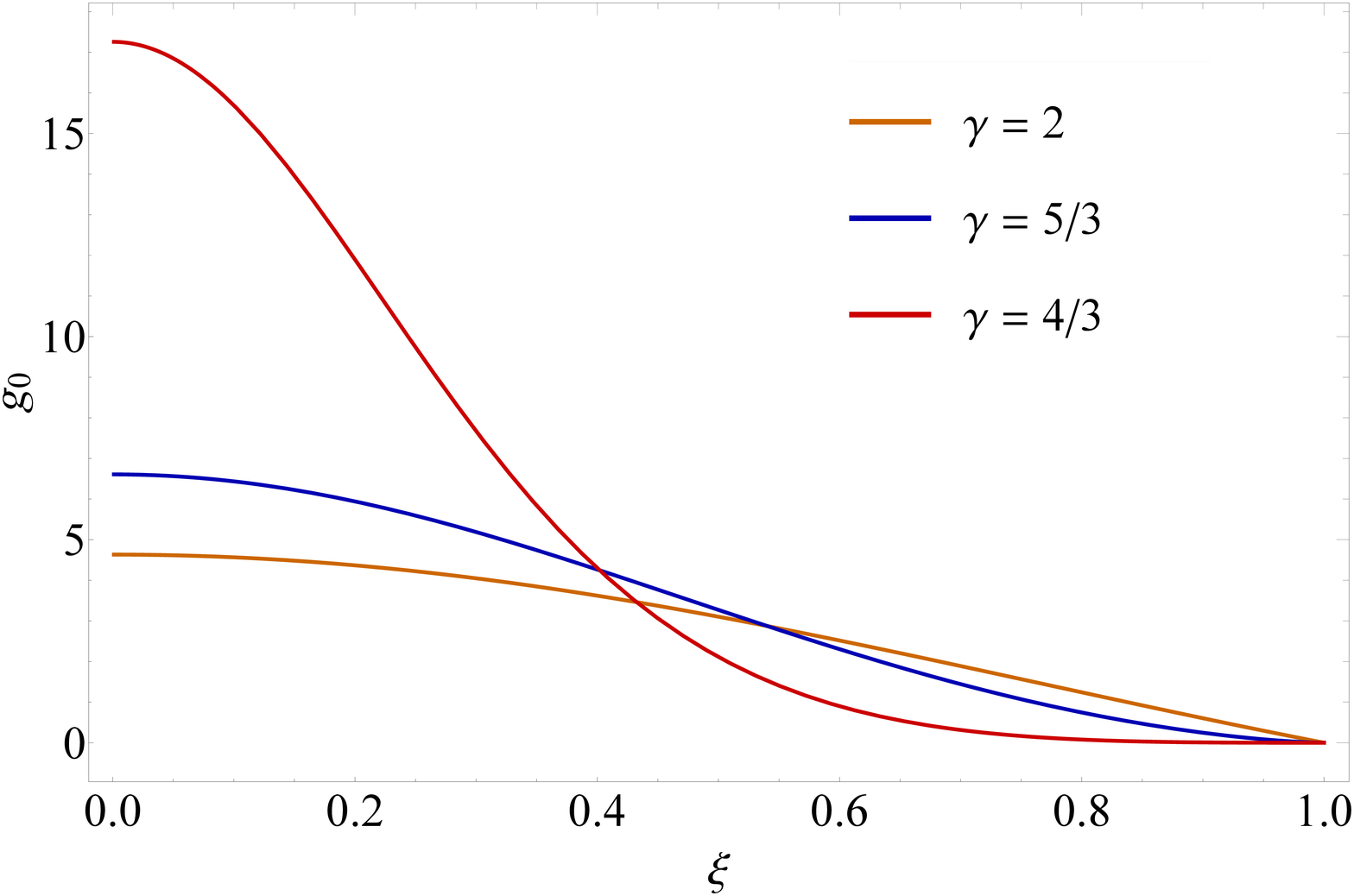} 
   \caption{Left: The functions $\lambda_0$, $g_0$, and $h_0$, being the dimensionless enclosed line mass, density, and pressure, as a function of the dimensionless radius for a polytropic index $\gamma = 5/3$; the density and pressure are normalized by their values along the axis of the cylinder, $g_0(0)$ and $h_0(0)$, respectively. Right: The solution for the density of the cylinder for the polytropic indices shown in the legend. These solutions have respective specific entropies of $K_0 = 0.0865$, 0.107, and 0.131. }
   \label{fig:lgh}
\end{figure}

The left panel of Figure \ref{fig:lgh} shows the solution for the dimensionless density, pressure, and line mass when $\gamma = 5/3$, for which $K_0 \simeq 0.107$; to aid in the visualization of these functions, here we normalized the dimensionless pressure and density $h_0$ and $g_0$ by their values at the origin, $g_0(0)$ and $h_0(0)$. The right panel of this figure gives the solution for the dimensionless density when $\gamma = 2$, 5/3, and 4/3, for which  $K_0 \simeq 0.0865$, 0.107, and 0.131 respectively.

\citet{ostriker64} described the properties of these solutions, and \citet{ostriker65} presented detailed tabulated values of the specific functions for a range of adiabatic indices. Here our primary concern is not with the hydrostatic solutions themselves, but with their response to imposed perturbations. We turn to the analysis of cylindrical-radial perturbations in the next section.

\section{Radial perturbations}
\label{sec:cylperts}
From the above definitions of the density and the pressure \eqref{rhodef}, the sound speed $c_{\rm s}$ is given by $c_{\rm s} \simeq \sqrt{G\Lambda}$, which yields the dimensionless sound-crossing time $\tau$ of the cylinder:

\begin{equation}
d\tau \equiv \frac{\sqrt{G\Lambda}}{H}dt \simeq \frac{c_{\rm s}dt}{H}.
\end{equation}
We now impose purely cylindrically radial perturbations (i.e., no variation along the axis of the cylinder or around the axis) on top of the hydrostatic cylinder that induce a cylindrical-radial velocity $v_{\rm s}$, which we parameterize by

\begin{equation}
v_{\rm s} = \sqrt{G\Lambda}f_{\rm s}(\xi,\tau). \label{vsdef}
\end{equation}
Here $f_{\rm s}$ is a function of $\xi = s/H(t)$, where $H(t)$ is the true stream radius (i.e., including the radial perturbations). The component of the velocity normal to the surface of the cylinder must be continuous across the surface in the comoving frame of the surface, which implies that the function $f_{\rm s}$ satisfies

\begin{equation}
f_{\rm s}(\xi = 1,\tau) = \frac{1}{\sqrt{G\Lambda}}\frac{\partial H}{\partial t}. \label{fsbc1}
\end{equation}
We thus see that if the motion of the surface is subsonic then the function $f_{\rm s}$ will be less than one. In this limit of subsonic surface motions we are justified in dropping nonlinear terms in the velocity from the fluid equations, and we can construct the linear response of the cylinder to such small-amplitude perturbations. We therefore expand the fluid density and pressure as

\begin{equation}
\rho = \frac{\Lambda}{4\pi H^2}\left\{g_0(\xi)+g_1(\xi,\tau)\right\}, \quad p = \frac{G\Lambda^2}{4\pi H^2}\left\{h_0(\xi)+h_1(\xi,\tau)\right\},
\end{equation}
where $g_1$ and $h_1$ are small perturbations to the density and pressure with respect to the unperturbed functions $g_0$ and $h_0$, respectively. Dropping nonlinear terms, we can take the Laplace transform of the continuity, radial momentum, and entropy equations to obtain three algebraic equations relating the perturbation to the pressure, the perturbation to the density, the cylindrical-radial velocity, and the perturbation to the surface velocity. These three equations can then be combined into a single, second-order differential equation for the cylindrical-radial velocity; to maintain the readability of the paper, the algebraic steps are deferred to Appendix \ref{sec:cylderivation}, and the result is

\begin{equation}
\sigma^2 \tilde{f}_{\rm s}+\frac{\partial}{\partial \xi}\left[\frac{\gamma h_0}{g_0}\frac{1}{g_0\xi}\frac{\partial}{\partial \xi}\left[g_0\xi\left(\tilde{V}\xi-\tilde{f}_{\rm s}\right)\right]\right]-2\left(\gamma-1\right)\tilde{V}\frac{1}{g_0}\frac{\partial h_0}{\partial \xi} = -g_0\left(\tilde{V}\xi-\tilde{f}_{\rm s}\right)+\sigma \delta f(\xi). \label{finaleq1}
\end{equation}
Here a tilde denotes the Laplace transform of a quantity, where the Laplace transform of $f_{\rm s}$ is

\begin{equation}
\tilde{f}_{\rm s}(\xi,\sigma) = \int_0^{\infty}f_{\rm s}(\xi,\tau)e^{-\sigma\tau}d\tau,
\end{equation}
and the quantity $V$ is

\begin{equation}
V \equiv \frac{1}{\sqrt{G\Lambda}}\frac{\partial H}{\partial t} = \frac{\partial}{\partial \tau}\ln H,
\end{equation}
and is the dimensionless velocity of the surface of the cylinder. Here we imposed an initial cylindrical-radial velocity $\delta f$ to seed the perturbations; modifying the approach to allow for an initial pressure excess that induces the motion, for example, is straightforward.

Equation \eqref{finaleq1} is our fundamental equation describing the radial oscillations of a polytropic cylinder, and to integrate it numerically for a given $\sigma$, $\tilde{V}$, and $\delta f$ we need two boundary conditions on the velocity. The first such boundary condition arises from the continuity of the normal component of the velocity across the surface, which yields Equation \eqref{fsbc1}, and shows that

\begin{equation}
\tilde{f}_{\rm s}(\xi = 1) = \tilde{V}.
\end{equation}
The second boundary condition comes from the requirement that the solution be expandable about the surface and non-trivial, which gives

\begin{equation}
\frac{d\tilde{f}_{\rm s}}{d\xi}\bigg{|}_{\xi = 1}  =  -\frac{1}{\gamma}\left\{\left(\sigma^2+\gamma-2\right)\tilde{V}-\sigma \delta f(1)\right\}. \label{bccyl}
\end{equation}
To derive this expression we used, from Equation \eqref{msurf}, that $g_0 \propto (1-\xi)^{1/(\gamma-1)}$ and $\gamma h_0/g_0 = \left(\gamma-1\right)\left(1-\xi\right)$. Finally, for a given $\sigma$ and initial velocity perturbation $\delta f$ and an arbitrarily-chosen $\tilde{V}$, integrating Equation \eqref{finaleq1} numerically from the surface will not necessarily result in a well-behaved solution near $\xi = 0$. However, we require that physical perturbations that maintain cylindrical symmetry do not displace the axis itself, and hence we require that $\tilde{f}_{\rm s}(\xi = 0) = 0$. For a given $\sigma$ (and initial velocity perturbation), there will be a unique $\tilde{V}$ that results in the solution for $\tilde{f}_{\rm s}$ satisfying this third boundary condition. The solution for the velocity of the surface itself is therefore constrained by requiring that solutions remain well-behaved along the axis of the cylinder.

\subsection{Eigenmodes}
There are special values of $\sigma \equiv \sigma_{\rm n}$, which in general are complex numbers, for which the solution for $\tilde{V}$ in Equation \eqref{finaleq1} diverges. In the vicinity of the point $\sigma_{\rm n}$, we can divide Equation \eqref{finaleq1} by $\tilde{V}$, define $\tilde{f}_{\rm n} \equiv \tilde{f}_{\rm s}/\tilde{V}$, and letting $\sigma \rightarrow \sigma_{\rm n}$, Equation \eqref{finaleq1} becomes

\begin{equation}
\sigma_{\rm n}^2 \tilde{f}_{\rm n}+\frac{\partial}{\partial \xi}\left[\frac{\gamma h_0}{g_0}\frac{1}{g_0\xi}\frac{\partial}{\partial \xi}\left[g_0\xi\left(\xi-\tilde{f}_{\rm n}\right)\right]\right]-2\left(\gamma-1\right)\frac{1}{g_0}\frac{\partial h_0}{\partial \xi} = -g_0\left(\xi-\tilde{f}_{\rm n}\right), \label{eigeneq1}
\end{equation}
with boundary conditions

\begin{equation}
\tilde{f}_{\rm n}(1) = 1, \quad \frac{d\tilde{f}_{\rm n}}{d\xi}\bigg{|}_{\xi = 1}  =  -\frac{1}{\gamma}\left(\sigma_{\rm n}^2+\gamma-2\right).
\end{equation}
The dependence on the specific perturbation drops out from both the differential equation and the boundary condition owing to the fact that $1/\tilde{V} \rightarrow 0$ as $\sigma \rightarrow \sigma_{\rm n}$ (i.e., the ratio $\delta f/\tilde{V}$ becomes much less than the rest of the operator equation in Equation \ref{finaleq1} as $\tilde{V} \rightarrow \infty$), and $\sigma_{\rm n}$ is determined by requiring that the solution for $\tilde{f}_{\rm n}$ be well-behaved (and equal to zero) at $\xi = 0$. In maintaining a finite $\tilde{f}_{\rm n}$ we are implicitly letting the eigenfrequencies describing the surface oscillations coincide with those of the eigenfunctions; assuming otherwise leads either to contradictions or trivial solutions, and hence the motions of the interior of the cylinder directly couple to the surface motions (i.e., one cannot have motion in the interior of the cylinder that does not impact the surface). The eigenvalues $\sigma_{\rm n}$ and the corresponding eigenfunctions are then independent of the specific perturbation that originated the movement of the cylinder. Also note that the eigenvalue equation \eqref{eigeneq1} depends only on $\sigma_{\rm n}^2$, and hence if we find any solution $\sigma_{\rm n}$, then $-\sigma_{\rm n}$ is also a solution with the same $\tilde{f}_{\rm n}$. 

\begin{figure*}[htbp] 
   \centering
    \includegraphics[width=0.495\textwidth]{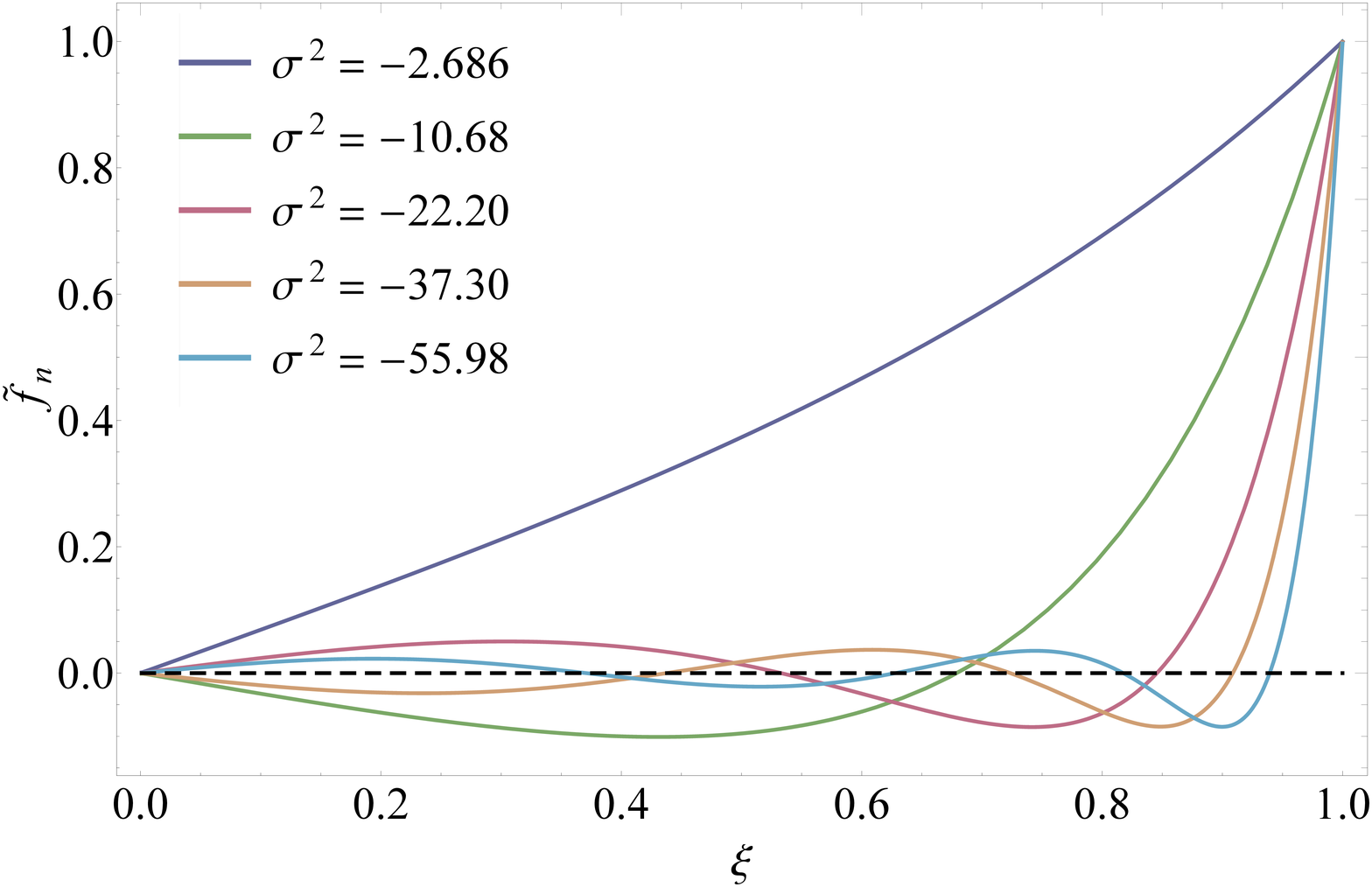} 
     \includegraphics[width=0.495\textwidth]{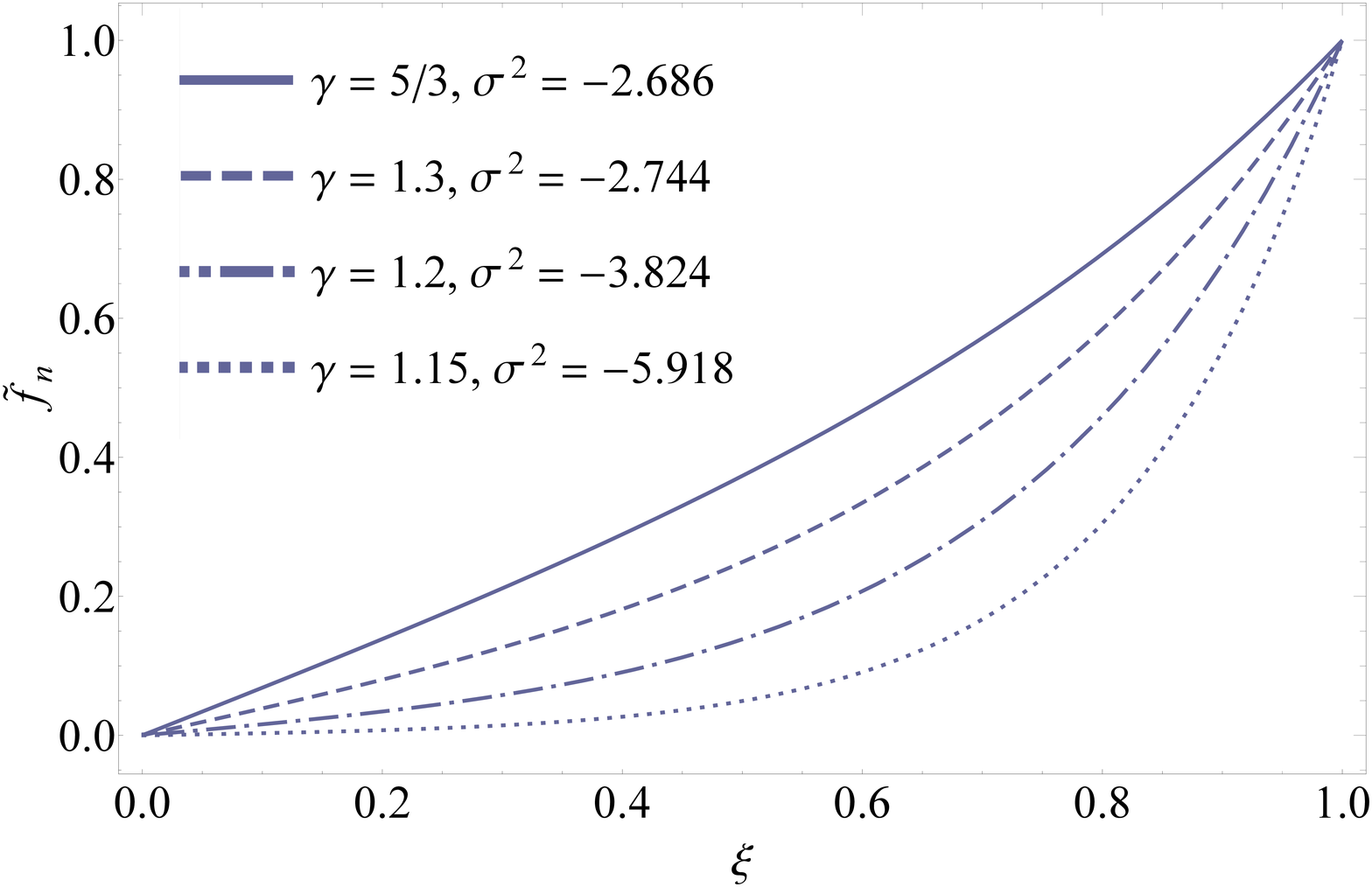} 
   \caption{Left: The first five eigenfunctions describing the cylindrical-radial velocity of a $\gamma = 5/3$ cylinder, where the legend gives the square of the eigenvalue appropriate to each mode. The fact that all of the eigenvalues are imaginary means that the solution is stable, and the cylinder oscillates in response to radial perturbations. The fundamental (lowest-order) mode has one zero crossing at the origin, and each successive mode has one more node than the previous one. The black, dashed line shows $\tilde{f}_{\rm s} = 0$ for reference. Right: The lowest-order mode for the adiabatic indices shown in the legend. As the adiabatic index decreases, the fundamental mode becomes increasingly nonlinear, implying that one needs many higher-order modes to reconstruct a linear (i.e., well-ordered) initial velocity perturbation. As the equation of state softens, the cylinder becomes unstable to chaotic motions in the outer, low-density region of the cylinder.}
   \label{fig:eigens}
\end{figure*}

The left panel of Figure \ref{fig:eigens} shows the first five eigenfunctions when $\gamma = 5/3$; the legend in this figure gives the square of the eigenvalue that corresponds to each mode, and demonstrates that all of these modes are characterized by purely imaginary eigenfrequencies. The lowest-order mode has a zero crossing only at the origin, and each higher-order mode has an additional node where the function intersects the line $\tilde{f}_{\rm s} = 0$, shown in this figure by the black, dashed line for reference. 

The eigenmodes offer a numerically-convenient means of constructing the response of the cylinder to an imposed perturbation, as we can write 

\begin{equation}
\tilde{V}\prod_{\rm n}\left(\sigma^2-\sigma_{\rm n}^2\right) = C,
\end{equation}
where $C$ is a holomorphic function in the complex plane. Note, however, that we can use the inverse Laplace transform, being

\begin{equation}
V(\tau) = \frac{1}{2\pi i}\int \tilde{V}(\sigma)e^{\sigma\tau}d\sigma,
\end{equation}
where the integral is taken over a line in the complex plane that extends from $-i\infty$ to $i\infty$ and is to the right of all the poles (the eigenvalues) of $\tilde{V}$ (see, e.g., \citealt{riley06}), to write the solution for $V$ as

\begin{equation}
V(\tau) = \frac{1}{2\pi i}\int \frac{C}{\prod_{\rm n}\left(\sigma^2-\sigma_{\rm n}^2\right)}e^{\sigma\tau}d\sigma = \frac{1}{2\pi i}\int\frac{C}{\left(\sigma^2-\sigma_{\rm j}^2\right)\prod_{\rm n \neq j}\left(\sigma^2-\sigma_{\rm n}^2\right)}e^{\sigma\tau}d\sigma
= \sum_{\rm n}\left(\frac{\partial}{\partial \sigma}\left[\frac{1}{\tilde{V}}\right]\bigg{|}_{\sigma = \sigma_{\rm n}}\right)^{-1}e^{\sigma_{\rm n}\tau}, \label{Vsigma}
\end{equation}
where in the last line we closed the contour integral in the left half of the complex plane and used the residue theorem. Thus, for a given initial perturbation, we need only calculate the quantity $1/\tilde{V}$ in the immediate vicinity of the eigenvalues $\sigma_{\rm n}$ to determine its derivative, and the response of the cylinder is then a sum over a discrete set of modes. If all of the eigenvalues are purely imaginary, then the hydrostatic configuration is stable and oscillates in response to an imposed perturbation. On the other hand, a mode with a positive real part implies that the solution is unstable, and perturbations grow exponentially rapidly.

For $\gamma = 5/3$, all of the eigenvalues are imaginary, and the response of such a polytropic cylinder to an imposed radial perturbation is to stably oscillate. Note, however, that if we let $\gamma = 1$, Equation \eqref{eigeneq1} appears to have the exact solution $\sigma_1 = 0$ and $\tilde{f}_{1} = \xi$, which suggests that isothermal cylinders are unstable to radial perturbations (recall that $\sigma$ refers to the eigenvalue of the \emph{velocity} of the surface of the cylinder, not its displacement, and hence $\sigma = 0$ corresponds to a constant surface velocity), which is the same result recovered by \citet{chandrasekhar53} and \citet{breysse14}. The adiabatic index of $\gamma = 1$ therefore appears analogous to $\gamma = 4/3$ for spherical polytropes.

The analogy is, however, not entirely accurate, as the physical behavior and manifestation of the instability between the two cases is fundamentally different: for the spherical case, as $\gamma$ decreases to 4/3 the lowest-order eigenfunction becomes more linear and the lowest eigenvalue approaches $\sigma = 0$. Thus, if we impose an initial velocity perturbation of $\delta f = -\xi$ on a polytropic star, then as $\gamma \rightarrow 4/3$ the response of the spherical polytrope is dominated by the lowest-order mode, and the star collapses inward on itself homologously and the velocity profile remains almost linearly radial with time (and is exactly linear when $\gamma = 4/3$). This behavior is ultimately due to a phase lag between the perturbation to the velocity and those to the pressure, density, and gravitational potential, as the latter are all proportional to the fluid displacement. The pressure therefore builds {in response} to the existence of the initial velocity field to withstand the radial collapse of the sphere. The gravitational potential, however, also builds precisely in sync with the pressure, which serves to accelerate the inward motion of the fluid. The value of $\gamma = 4/3$ thus represents the critical adiabatic index at which the pressure cannot build vigorously enough to overcome the destabilizing nature of self-gravity, and the spherical polytrope continues to collapse in on itself at a constant velocity. The constancy of the velocity arises from the competition between the simultaneously rising pressure and gravitational potential.

On the other hand, the nature of the instability in the cylindrical case is completely different, as the lowest order mode becomes increasingly \emph{nonlinear} as the gas becomes more isothermal. This behavior is shown in the right panel of Figure \ref{fig:eigens}, which gives the fundamental mode for the adiabatic indices shown in the legend (the square of the eigenvalue is also given in the legend). If we therefore impose the same initial velocity profile of $\delta f = -\xi$ to a nearly isothermal cylinder, then instead of the response being dominated by the lowest-order mode, we instead need many higher order modes to accurately reconstruct the solution near the surface. Thus, instead of collapsing homologously inward, the cylinder responds to an initial radial infall by transferring power to higher-order modes, and the outer, low-density regions of the nearly-isothermal cylinder oscillate violently and stochastically. As we reach the threshold of $\gamma = 1$, the unstable mode appears to emerge as the new lowest-order mode with $\sigma^2 = 0$. 

However, the case of $\gamma = 1$ cannot be self-consistently analyzed with our set of variables, as the mass of the cylinder is infinitely concentrated along the axis and the radius of the cylinder is not a well defined quantity in the isothermal limit. Instead, in this case we would have to work with variables normalized by the quantities along the axis of the cylinder and the radius normalized by the central scale height, being $\sqrt{p_{\rm c}/(4\pi G\rho_{\rm c}^2)}$, where $p_{\rm c}$ and $\rho_{\rm c}$ are the pressure and density along the axis, respectively. In terms of these variables, as $\gamma \rightarrow 1$, the surface of the cylinder grows to an arbitrarily large number of scale heights. Thus, at $\gamma = 1$, the cylinder becomes infinite in extent, and the set of eigenmodes goes from a set of discrete points to a dense continuum. The function $\tilde{f}_{\rm s} = \xi$ is also not the solution at $\sigma^2 = 0$, as there is no longer a surface at which to apply a regularity boundary condition; instead, the boundary condition on the eigenmodes is such that there are no inward-propagating waves as we move to larger radii, as the background state now possesses what is effectively an outflow boundary at large radii. Thus, while $\gamma = 1$ possesses the solution $\sigma^2 = 0$, it also possesses an infinite number of solutions with $\sigma^2 = \epsilon$ and $\epsilon \rightarrow 0$.

With these points in mind, we now move on to the analysis of general (i.e., not necessarily restricted to the radial direction) perturbations of an adiabatic, self-gravitating cylinder and analyze the eigenmodes of these cylinders. We note that our analytical approach and numerical methodology for calculating the eigenmodes differs from what is commonly presented in, for example, texts on stellar oscillations (e.g., \citealt{cox80}). To verify the validity of our methods, we also derived the eigenmode equations that describe the oscillations of a polytropic sphere and recovered identical results to those presented in \citet{lee86} for the eigenfrequencies of the $f$, $p_1$, $p_2$, etc.~modes of a $\gamma = 5/3$ polytrope. Because they may be useful in other contexts and we have not encountered them elsewhere, we give these equations in Appendix \ref{sec:spherical} for reference.

\section{Eigenmode equations for Non-radial perturbations}
\label{sec:general}
The previous subsection assumed that perturbations were purely in the cylindrical-radial direction. Here we allow the perturbations to also be along the axis of the cylinder (the $z$-direction) and around the axis of the cylinder (the $\varphi$-direction). As above, we non-dimensionalize the fluid quantities by defining

\begin{equation}
v_{\rm s, z, \varphi} = \sqrt{G\Lambda}f_{\rm s,z, \varphi}(\xi,\eta,\varphi,\tau), \quad \rho = \frac{\Lambda}{4\pi H^2}\left\{g_0(\xi)+g_1(\xi,\eta,\varphi,\tau)\right\}, \quad p = \frac{G\Lambda^2}{4\pi H^2}\left\{h_0(\xi)+h_1(\xi,\eta,\varphi,\tau)\right\}
\end{equation}
where

\begin{equation}
\xi = \frac{s}{H(z,\varphi,t)}, \quad \eta = \frac{z}{H(z,\varphi,t)}, \quad d\tau = \frac{\sqrt{G\Lambda}}{H(z,\varphi,t)}dt. \label{dtaudt}
\end{equation}
Here $H$ is the true location of the surface of the cylinder, i.e., including the perturbations induced by the fluid motion, and the cylindrical nature of the unperturbed solution demands that $\partial H/\partial z \ll 1$ and $\partial H/\partial \varphi \ll H$ in the perturbative limit. As for the previous section, the subsonic nature of the perturbations also implies that $\partial H/\partial t \ll \sqrt{G\Lambda}$. We now must also explicitly account for the perturbations to the gravitational potential of the cylinder, $\Phi$, and the surface of the cylinder, $H$, which we parameterize by

\begin{equation}
\Phi = G\Lambda\left\{j_0(\xi)+j_1(\xi,\eta,\varphi,\tau)\right\}, \quad H(z,\varphi,t) = H_0\left\{1+\zeta\left(\eta,\varphi,\tau\right)\right\}.
\end{equation}
Here $H_0$ is the unperturbed radius of the cylinder, and $\zeta$ is the dimensionless perturbation to the surface. Unlike the case of radial perturbations, for which the perturbation to the surface velocity is the only quantity that has physical meaning\footnote{By this we mean that we are always free to rescale the units of the problem, and hence when the oscillations are purely radial the fundamental physical quantity is the velocity of the surface of the cylinder, not its position; this is apparent from the fact that in Section \ref{sec:cylperts} we were able to solve for the perturbation to the velocity and never had to introduce the perturbation to the surface itself, and hence the only restriction that ensures the quasi-hydrostatic nature of the solution is that the surface velocity be small in comparison to the sound speed.}, here the presence of deviations along and around the axis of the cylinder require that we account for the angular and linear variations of the surface itself. We also normalize the unperturbed gravitational potential by $j_0(1) = 0$; the normalization does not, of course, affect the properties of the solutions, but it does simplify the appearance of the equations.

We can now insert the above definitions into the fluid equations and the Poisson equation, take their Fourier and Laplace transforms, and derive the linearized perturbation equations and the eigenvalue equations; we defer the details of the calculations to Appendix \ref{sec:derivation} and present only the final results here, being the following set of two equations for the functions $\tilde{F}_{\rm s} \equiv \sigma\tilde{f}_{\rm s}$ and $\tilde{j}_1$:

\begin{equation}
\tilde{F}_{\rm s}+\frac{\partial}{\partial \xi}\left[ \frac{\frac{1}{g_0\xi}\frac{\partial}{\partial \xi}\left[g_0\xi\left(\sigma^2\tilde{\zeta}\xi-\tilde{F}_{\rm s}\right)\right]\frac{\gamma h_0}{g_0}+\sigma^2\left(\tilde{j}_{1}-\frac{2\gamma h_0}{g_0}\tilde{\zeta}\right)}{\sigma^2+\left(k^2+\frac{m^2}{\xi^2}\right)\frac{\gamma h_0}{g_0}}\right] = \delta\tilde{f}_{\rm s} \label{Fex1}
\end{equation}
\begin{equation}
\frac{1}{\xi}\frac{\partial}{\partial \xi}\left[\xi\frac{\partial \tilde{j}_{1}}{\partial \xi}\right]-\left(k^2+\frac{m^2}{\xi^2}\right)\left(\tilde{j}_{1}-\tilde{\zeta}\xi\frac{\partial j_0}{\partial \xi}\right) = g_0\frac{\frac{1}{g_0\xi}\frac{\partial}{\partial\xi}\left[g_0\xi\left(\sigma^2\tilde{\zeta}\xi-\tilde{F}_{\rm s}\right)\right]-\left(k^2+\frac{m^2}{\xi^2}\right)\left(\tilde{j}_{1}-\frac{2\gamma h_0}{g_0}\tilde{\zeta}\right)}{\sigma^2+\left(k^2+\frac{m^2}{\xi^2}\right)\frac{\gamma h_0}{g_0}}. \label{jex1}
\end{equation}
The first of these is the radial momentum equation and we let there be an initial cylindrical-radial velocity perturbation $\delta f_{\rm s}(\xi,\eta,\varphi)$ that drives the motion of the cylinder, while the second is the Poisson equation. Tildes denote Laplace and Fourier-transformed quantities so that, for example, $\tilde{f}_{\rm s}$ is given by

\begin{equation}
\tilde{f}_{\rm s}(\xi,\sigma,k,m) = \int_0^{\infty}\int_{-\infty}^{\infty}\int_0^{2\pi}f_{\rm s}(\xi,\tau,\eta,\varphi)e^{-\sigma\tau-ik\eta-im\varphi}d\varphi \, d\eta \, d\tau,
\end{equation}
and similarly for $\tilde{j}_1$. To ensure the periodicity of the solution around the axis of the cylinder, $m$ is restricted to positive and negative integers, while $k$ is a continuous variable. It is also apparent from Equations \eqref{Fex1} and \eqref{jex1} that the signs of $m$ and $k$ do not affect the solution, and we can restrict our analysis to positive $m$ and $k$. 

Analogously to what was done in Section \ref{sec:cylperts} for purely radial perturbations, the boundary conditions at the surface can be determined by Taylor expanding the functions $\tilde{F}_{\rm s}$ and $\tilde{j}_{1}$ about $\xi = 1$ and requiring that the normal component of the velocity be continuous across the surface of the cylinder; writing the leading-order series expansion of the gravitational potential about the surface as $\tilde{j}_{1}/\tilde{\zeta}= \tilde{J}+\tilde{J}'(1-\xi)$, this gives 

\begin{equation}
\tilde{F}_{\rm s}(\xi = 1) = \sigma^2\tilde{\zeta}, \quad \frac{d\tilde{F}_{\rm s}}{d\xi}\bigg{|}_{\xi = 1} = -\frac{\sigma^2}{\gamma}\tilde{\zeta}\left(\sigma^2+\gamma-2+\left(\gamma-1\right)\left(k^2+m^2\right)\frac{1}{\sigma^2}\tilde{J}-\tilde{J}'-\frac{1}{\tilde{\zeta}}\delta\tilde{f}(\xi = 1)\right). \label{bceigens}
\end{equation}
We can obtain a second relation between $\tilde{J}$ and $\tilde{J}'$ by noting that, outside the surface of the cylinder ($\xi > 1$), the density is exactly zero and Equation \eqref{jex1} can be written as Laplace's equation in cylindrical coordinates for the quantity $\Delta \tilde{j}_{1} \equiv \tilde{j}_{1}-\tilde{\zeta}\xi\partial j_0/\partial \xi$. There are two independent solutions to this equation, one of which diverges at large $\xi$ while the other decays; discarding the growing solution as unphysical, as small perturbations to the fluid should not result in diverging corrections to the gravitational potential, we find that for all $\xi > 0$ the potential satisfies

\begin{equation}
\Delta \tilde{j}_{1} \propto H^{(1)}_{\rm m}(ik\xi).
\end{equation}
$H^{(1)}_{\rm m}$ is a Hankel function of the first kind, which in the limit that $k \rightarrow 0$ approaches $\xi^{-|m|}$. To ensure that the gravitational field remains finite, the potential itself and the derivative of the potential must be continuous across the surface of the cylinder, which shows that $\tilde{J}$ and $\tilde{J}'$ are related by

\begin{equation}
\tilde{J}'= -\frac{\tilde{J}-1}{H_{\rm m}^{(1)}(ik)}\frac{\partial}{\partial \xi}\left[H_{\rm m}^{(1)}\left(ik\xi\right)\right]\bigg{|}_{\xi = 1}. \label{Jbc}
\end{equation}
Equations \eqref{bceigens} and \eqref{Jbc} can be used to integrate Equations \eqref{Fex1} and \eqref{jex1} for a given $\sigma$, $\tilde{J}$, and $\tilde{\zeta}$. For a chosen $\sigma$, only for a special combination of $\tilde{J}$ and $\tilde{\zeta}$ will the solutions remain regular at the origin. As for the purely radial perturbations, the displacement of the cylinder is recovered by ensuring that the solutions are well-behaved along the axis. In this case, we also recover the perturbation to the gravitational potential at the surface of the cylinder, $\tilde{J}$, which is constrained by the non-divergence of the gravitational field at the axis of the cylinder.

The eigenmodes are obtained when the solution for $\tilde{\zeta}$ has a pole in the complex plane, and performing the same steps as in Section \ref{sec:cylperts} and defining $\tilde{F}_{\rm n} \equiv \tilde{F}_{\rm s}/\tilde{\zeta}$ and $\tilde{j}_{\rm n} = \tilde{j}_1/\tilde{\zeta}$, we find that the eigenmodes satisfy the following two differential equations:

\begin{equation}
\tilde{F}_{\rm n}+\frac{\partial}{\partial \xi}\left[ \frac{\frac{1}{g_0\xi}\frac{\partial}{\partial \xi}\left[g_0\xi\left(\sigma_{\rm n}^2\xi-\tilde{F}_{\rm n}\right)\right]\frac{\gamma h_0}{g_0}+\sigma_{\rm n}^2\left(\tilde{j}_{\rm n}-\frac{2\gamma h_0}{g_0}\right)}{\sigma_{\rm n}^2+\left(k^2+\frac{m^2}{\xi^2}\right)\frac{\gamma h_0}{g_0}}\right] = 0, \label{Feigentot}
\end{equation}
\begin{equation}
\frac{1}{\xi}\frac{\partial}{\partial \xi}\left[\xi\frac{\partial \tilde{j}_{\rm n}}{\partial \xi}\right]-\left(k^2+\frac{m^2}{\xi^2}\right)\left(\tilde{j}_{\rm n}-\xi\frac{\partial j_0}{\partial \xi}\right) = g_0\frac{\frac{1}{g_0\xi}\frac{\partial}{\partial\xi}\left[g_0\xi\left(\sigma_{\rm n}^2\xi-\tilde{F}_{\rm n}\right)\right]-\left(k^2+\frac{m^2}{\xi^2}\right)\left(\tilde{j}_{\rm n}-\frac{2\gamma h_0}{g_0}\right)}{\sigma_{\rm n}^2+\left(k^2+\frac{m^2}{\xi^2}\right)\frac{\gamma h_0}{g_0}}, \label{jeigentot}
\end{equation}
and the boundary conditions

\begin{equation}
\tilde{F}_{\rm n} = \sigma_{\rm n}^2, \quad \frac{d\tilde{F}_{\rm n}}{d\xi}\bigg{|}_{\xi = 1} = -\frac{\sigma_{\rm n}^2}{\gamma}\left(\sigma_{\rm n}^2+\gamma-2+\left(\gamma-1\right)\left(k^2+m^2\right)\frac{1}{\sigma_{\rm n}^2}\left(J_{\rm n}+1\right)+\frac{J_{\rm n}}{H_{\rm m}^{(1)}(ik)}\frac{\partial}{\partial \xi}\left[H_{\rm m}^{(1)}(ik\xi)\right]\bigg{|}_{\xi = 1}\right),
\end{equation}
\begin{equation*}
\tilde{j}_{\rm n}(1) = J_{\rm n}, \quad \frac{d\tilde{j}_{\rm n}}{d\xi}\bigg{|}_{\xi = 1} = \frac{J_{\rm n}}{H_{\rm m}^{(1)}(ik)}\frac{\partial}{\partial \xi}\left[H_{\rm m}^{(1)}(ik\xi)\right]\bigg{|}_{\xi = 1}.
\end{equation*}
Here $J_{\rm n} \equiv \tilde{J}(\sigma_{\rm n})$. We can also derive an expression for the eigenfunction appropriate to the $z$-component of the velocity; see Equation \eqref{eigenfz}. The eigenvalues $\sigma_{\rm n}$ and $J_{\rm n}$ are recovered by requiring that the functions $\tilde{F}_{\rm n}$ and $\tilde{j}_{\rm n}$ remain well-behaved near the origin, and in particular that they retain finite values and derivatives at $\xi = 0$. We can determine these eigenvalues numerically by first picking a value of $\sigma_{\rm n}$ and iteratively calculating the $J_{\rm n}$ that maintains the regularity of the velocity at the origin: for the chosen $\sigma_{\rm n}$, we start with a guess for $J_{\rm n}$, numerically integrate Equations \eqref{Feigentot} and \eqref{jeigentot} inward from the surface, and calculate the value of $\tilde{F}_{\rm n}(\xi \simeq 0)$; we then perturb the guess for $J_{\rm n}$, calculate the new value of $\tilde{F}_{\rm n}(\xi \simeq 0)$, and use the difference between the new and old values of $\tilde{F}_{\rm n}(\xi \simeq 0)$ to inform our new guess for $J_{\rm n}$ that will better satisfy the boundary condition on $\tilde{F}_{\rm n}$ at the origin. We then continue to iteratively perturb our choice of $J_{\rm n}$ until the function $\tilde{F}_{\rm n}$ satisfies the boundary condition near the origin. For the same $\sigma_{\rm n}$ we can then perform precisely the same procedure to find the $J_{\rm n}$ that maintains the regularity of the gravitational potential near the origin. Only for special values of $\sigma_{\rm n}$ will the two $J_{\rm n}$ coincide, and these are then the eigenvalues $\sigma_{\rm n}$ and $J_{\rm n}$ that simultaneously satisfy the boundary conditions at the surface and the axis of the cylinder. 

The quantities $J_{\rm n}$ and $\sigma_{\rm n}$ could be complex numbers, implying that there are actually four unknowns -- the real and imaginary components of $\sigma_{\rm n}$ and $J_{\rm n}$ -- that are determined by simultaneously satisfying four boundary conditions -- the real and imaginary components of $\tilde{F}_{\rm n}$ and $\tilde{j}_{\rm n}$ being well behaved near the origin. However, note that Equations \eqref{Feigentot} and \eqref{jeigentot} depend only on $\sigma_{\rm n}^2$, and through a suitable redefinition of the eigenfunctions it can be shown that Equations \eqref{Feigentot} and \eqref{jeigentot} can be written in the form of a Hermitian operator equation in $\sigma_{\rm n}^2$. The eignvalues $\sigma_{\rm n}^2$ are therefore purely real, and -- as for the purely radial perturbations -- the dependence only on $\sigma_{\rm n}^2$ implies that if $\sigma_{\rm n}$ is a solution, then so is $-\sigma_{\rm n}$. 

When $\sigma_{\rm n}^2$ is real, this method of calculating the modes also offers a convenient way of visualizing the solutions: as we augment the value of $\sigma_{\rm n}$, the $J_{\rm n}$ that satisfies the boundary condition on the velocity or the gravitational potential adopts a new value, and the solutions for $J_{\rm n}(\sigma_{\rm n})$ therefore trace out curves in $J_{\rm n}-\sigma_{\rm n}$ space. The points where these two curves intersect then delimit the eigenvalues, and we can -- by calculating each $J_{\rm n}$ over a fairly wide and somewhat coarse range of $\sigma_{\rm n}$ -- visually inspect the intersection points and recover the eigenvalues. 

In the next section we use this method to calculate the eigenvalues corresponding to $m = 0$ and a range of $k$, and we show that there is an unstable mode that exists below a critical value of $k$ that describes the collapse of the cylinder along its axis. For the remainder of this paper we do not analyze modes with $m > 0$, as we have not found any unstable modes that characterize such perturbations\footnote{Though we note that $m = 1$, $k = 0$ has the exact solution $\sigma_{\rm n}^2 = 0$; since $m = 1$ modes yield displacements of the axis of the cylinder, this mode describes uniform translation of the entire cylinder in the radial direction and is another manifestation of the Galilean invariance of the fluid equations.}. This finding is consistent with that of \citet{hunter98}, who showed that -- in the limit that the cylinder has a constant density -- surface modes generated by a confining, ambient gas corresponding to $m > 0$ are stable when the pressure of the filament is much larger than that of the ambient medium. This stability feature was also found by \citet{aung19}, who investigated the presence and evolution of gravitational instability alongside the Kelvin-Helmoltz instability (we note that, as compared to these previous works, we are studying the ``body mode'' regime in which there is no confining pressure, i.e., the motions and oscillations of the fluid cylinder are due purely to the pressure and gravitational field of the cylinder itself and not to any external medium). For definiteness we focus primarily on gas-pressure dominated cylinders with $\gamma = 5/3$, though we plot the unstable eigenfrequency as a function of $k$ for various adiabatic indices. 

\section{Gravitational Instability of \MakeLowercase{\emph{m} = 0} modes}
\label{sec:m0modes}
\begin{figure}[htbp] 
   \centering
   \includegraphics[width=0.495\textwidth]{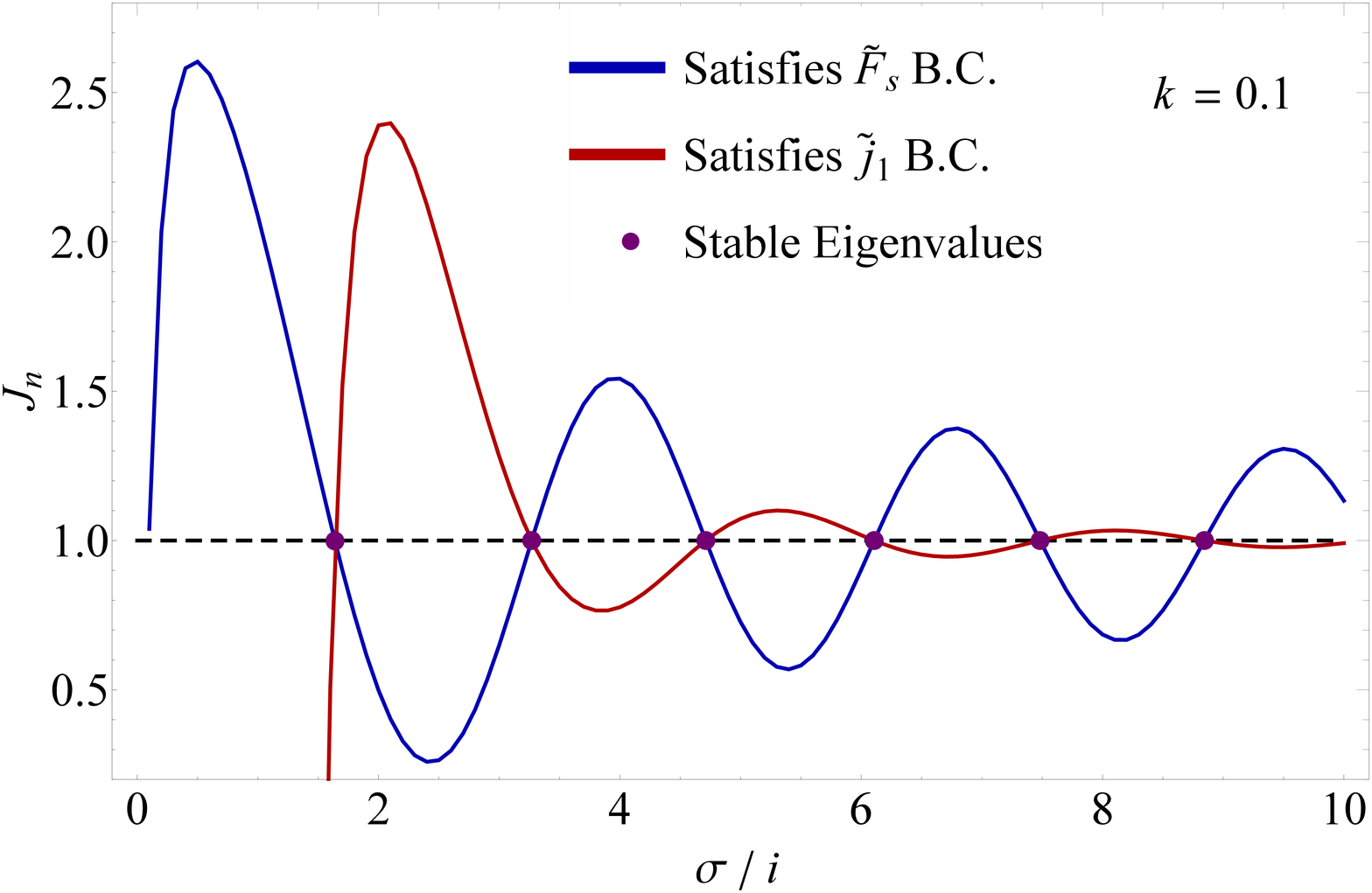} 
    \includegraphics[width=0.495\textwidth]{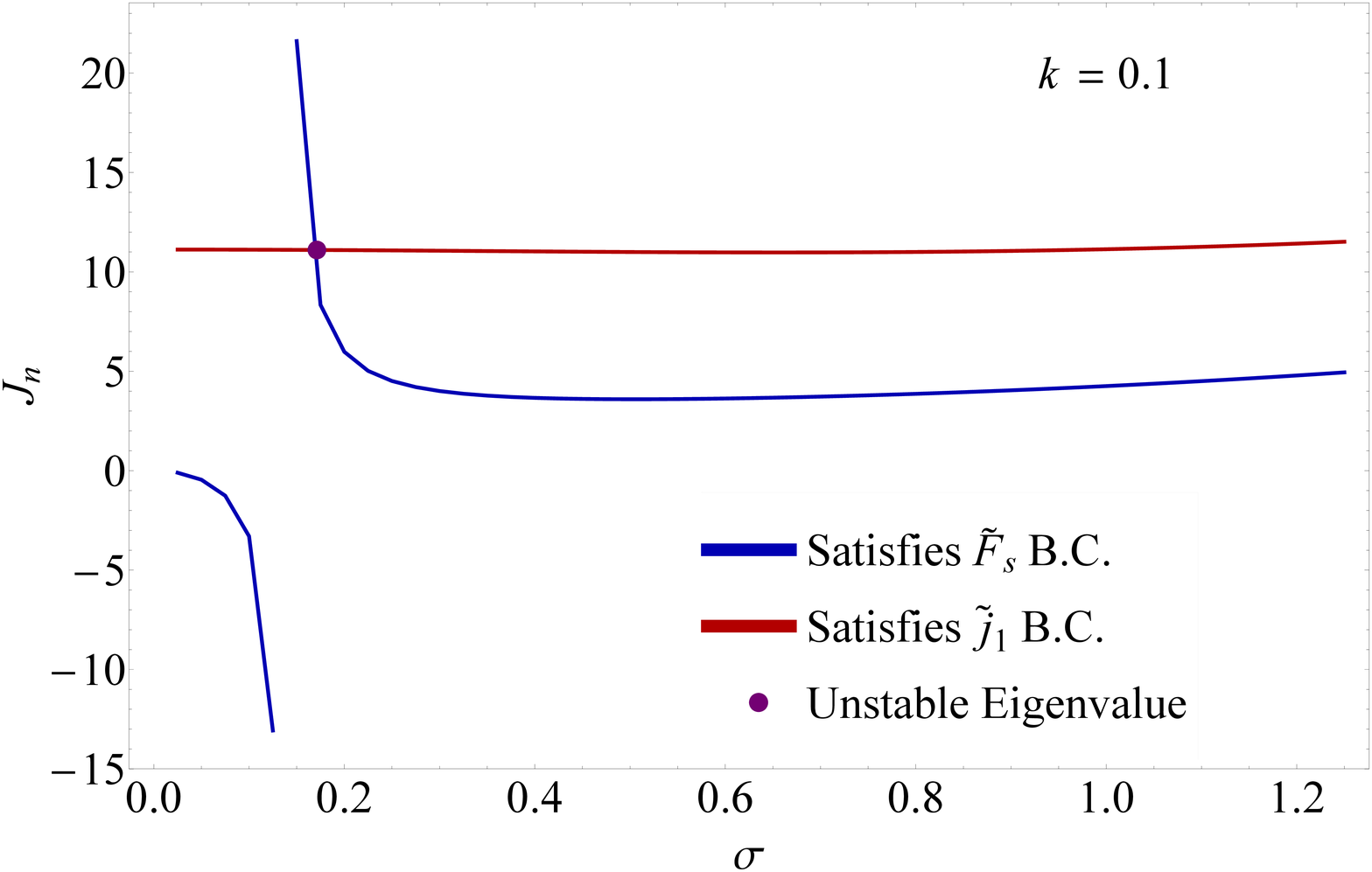} 
       \includegraphics[width=0.495\textwidth]{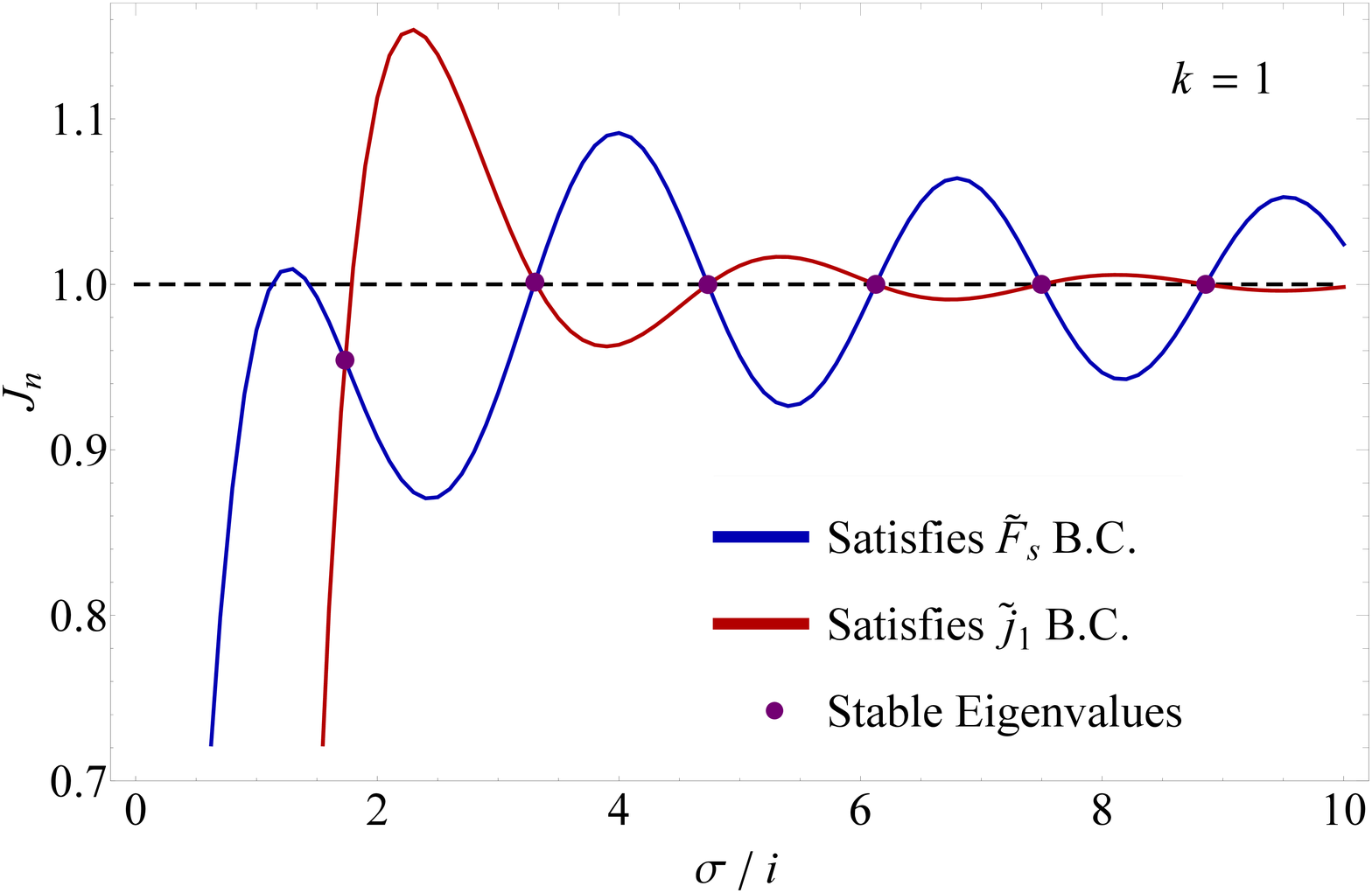} 
    \includegraphics[width=0.495\textwidth]{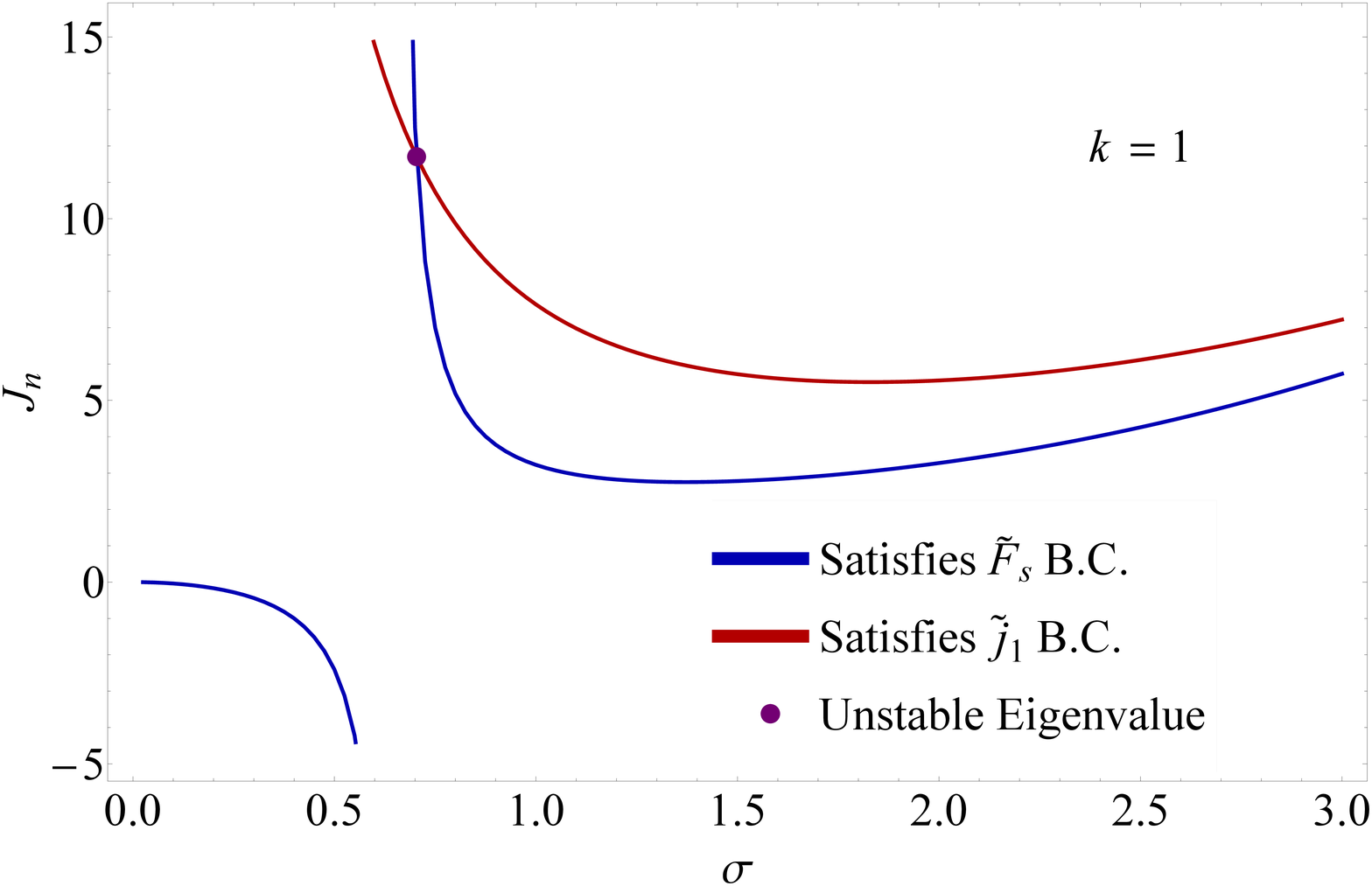} 
    \includegraphics[width=0.495\textwidth]{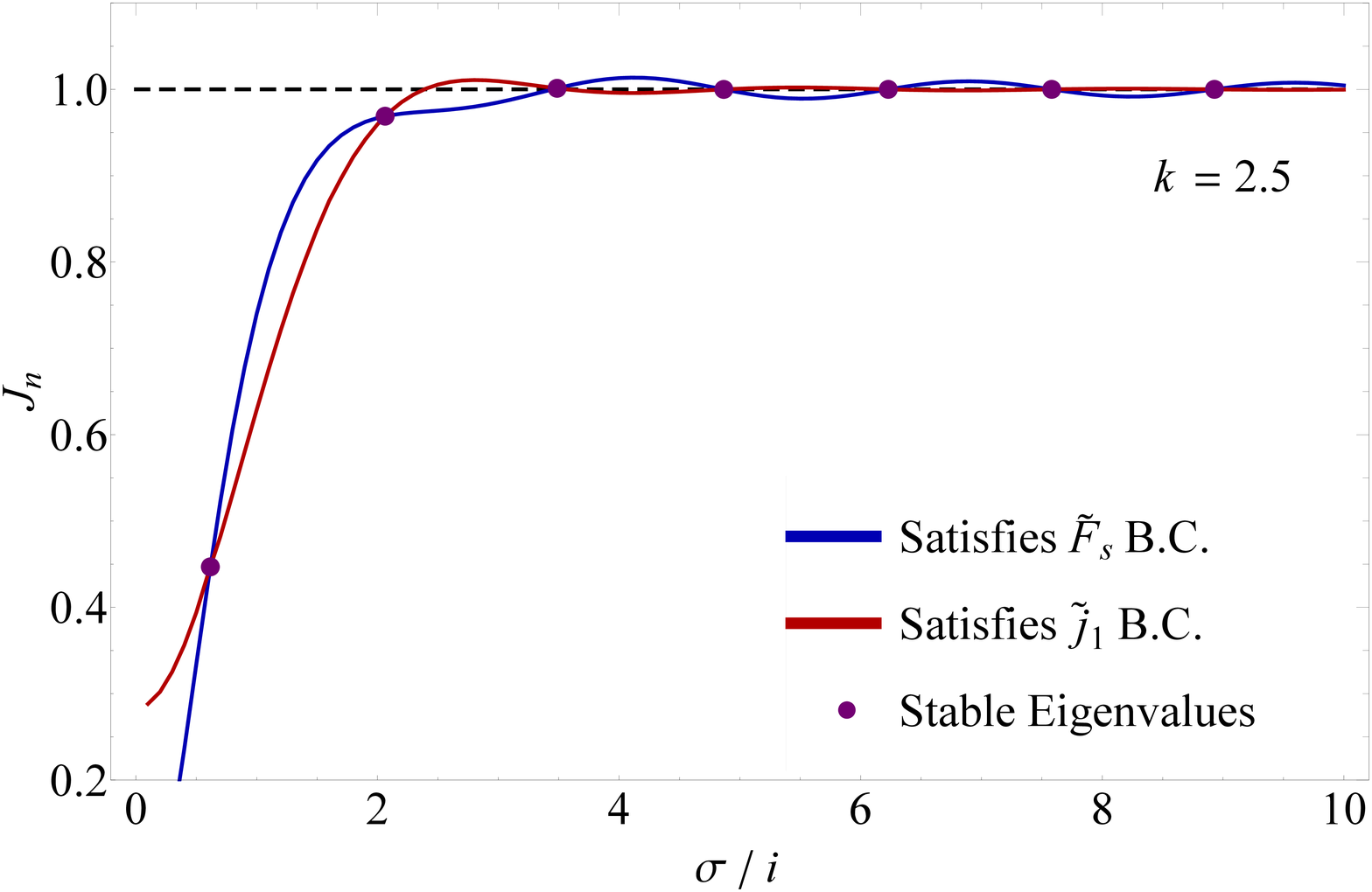} 
    \includegraphics[width=0.495\textwidth]{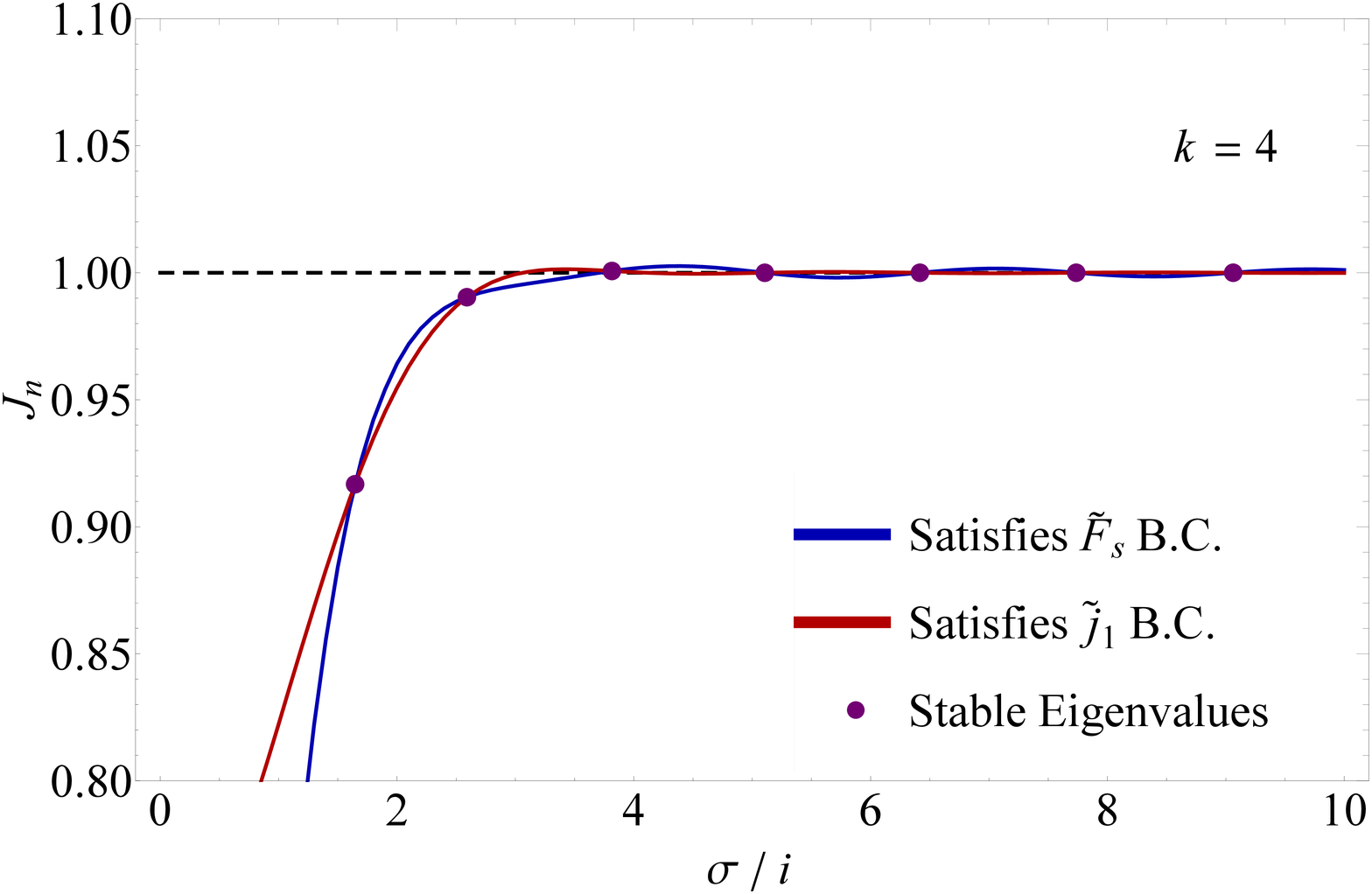} 
   \caption{The blue curves denote the values of $J_{\rm n}$ as functions of $\sigma$ that solve the eigenvalue equations (cf.~Equations \ref{Fex1} and \ref{jex1}) and the boundary condition $\tilde{F}_{\rm s}(\xi \simeq 0) = 0$, while red curves satisfy $\partial \tilde{j}_1/\partial \xi (\xi \simeq 0) = 0$. The value of the wavenumber $k$ is shown in the top-right of each panel; the top and middle left panels plot the $J_{\rm n}$ as functions of imaginary $\sigma$, the top and middle right as functions of real $\sigma$, and the bottom left and right as functions of imaginary $\sigma$. The intersections of the two curves (purple points) are the combinations of $J_{\rm n}$ and $\sigma$ that simultaneously satisfy both regularity conditions at the origin, and therefore are the eigenvalues, and the dashed lines show $J_{\rm n} = 1$ for reference. All of these solutions are for gas-pressure dominated cylinders with $\gamma = 5/3$. For $k = 0.1$ and $k = 1$ there are six stable eigenvalues and one unstable eigenvalue with $|\sigma| < 10$, while $k = 2.5$ and $k = 4$ possess seven stable eigenvalues (we did not plot $J_{\rm n}$ as functions of real $\sigma$ for $k = 2.5$ and $k = 4$ because there are no unstable modes for these $k$). The fact that each $k$ possesses the same total number of eigenvalues suggests that the lowest-frequency mode transitions from being unstable to stable above a critical wavenumber.}
   \label{fig:kp1_k1_k2}
\end{figure}

Figure \ref{fig:kp1_k1_k2} shows the $J_{\rm n}$ curves as functions of $\sigma$ for $\gamma = 5/3$ and the wavenumber $k$ shown in the top right of each panel; the top-right and middle-right panels show $J_{\rm n}$ as functions of real $\sigma$, whereas the remainder are functions of imaginary $\sigma$. The blue curves are the solutions that satisfy the boundary condition on the velocity at $\xi \simeq 0$, while the red curves satisfy the boundary condition on the gravitational potential, and the black, dashed lines show $J_{\rm n} = 1$ for reference. The intersections (purple points) are the eigenvalues, which simultaneously satisfy the regularity conditions on both the velocity and the gravitational potential. This figure demonstrates that perturbations with $k = 0.1$ and $k = 1$ are characterized by six stable eigenvalues and one unstable eigenvalue, while $k = 2.5$ and $k = 4$ possess seven oscillatory and stable eigenvalues with $|\sigma| < 10$ (we do not plot the $J_{\rm n}$ as functions of real $\sigma$ for $k = 2.5$ and $k = 4$ as there are no real eigenvalues). The fact that there are seven total modes for all $k$ suggests that the mode characterized by the smallest value of $|\sigma|$ transitions from being unstable to stable above a critical value of the wavenumber $k$.  We refer to the smallest mode by $\sigma_{\rm u}$ and, in line with the notation used in stellar oscillation theory, we denote the first stable mode by the $f$ mode, the second stable mode by $p_1$, the third stable mode by $p_2$, etc.; the specific eigenvalues are given in Table \ref{tab:1}. 

\begin{figure}[htbp] 
   \centering
   \includegraphics[width=0.995\textwidth]{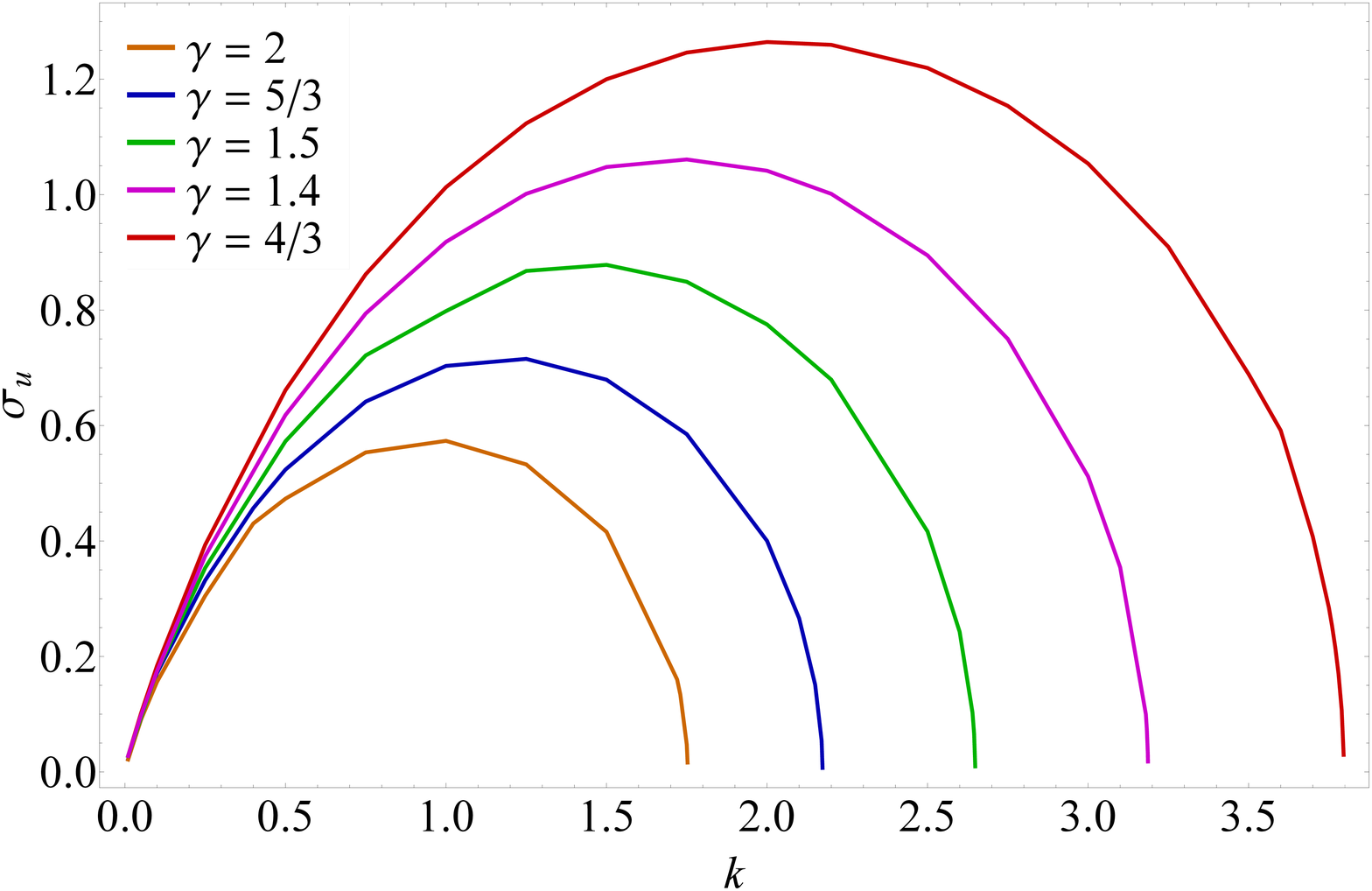} 
   \caption{The real part of the unstable eigenmode $\sigma_{\rm u}$ as a function of wavenumber $k$ for the adiabatic indices shown in the legend. The unstable mode is characterized by a small real part for $k \ll 1$, reaches a maximum value between 0.5 -- 1.2 at wavenumbers of order unity, and equals zero at a critical wavenumber $k_{\rm crit} \sim few$. For this range of $k$ the imaginary component of $\sigma_{\rm u}$ is zero, while above this range the real part is zero; the mode $\sigma_{\rm u}$ therefore goes from being unstable in the limit of $k \le k_{\rm crit}$ to stable for $k > k_{\rm crit}$. At $k = k_{\rm crit}$ the instability grows linearly with time. The maximum growth rate, $\sigma_{\rm max}$, the wavenumber at which the maximum occurs, $k_{\rm max}$, and the critical wavenumber above which the modes are stable, $k_{\rm crit}$, are given in Table \ref{tab:2}. }
   \label{fig:kunstable}
\end{figure}

To substantiate the notion that the smallest eigenvalue $\sigma_{\rm u}$ (as measured by $|\sigma| = \sqrt{\sigma^*\sigma}$) transitions from being unstable to stable above a critical $k$, Figure \ref{fig:kunstable} shows the real part of the unstable mode as a function of $k$ for the adiabatic indices in the legend. We see that, for $k \ll 1$, the real part of the eigenvalue is small and the growth rate of the corresponding instability is slow. As $k$ increases the real part of $\sigma_{\rm u}$ increases and reaches a maximum value, $\sigma_{\rm max}$, at a wavenumber $k_{\rm max}$ that is of the order unity. Beyond this wavenumber the real part decreases and equals zero at a critical wavenumber, $k_{\rm crit}$, and beyond this wavenumber the eigenvalue $\sigma_{\rm u}$ is purely imaginary and describes stable oscillations. The approximate values of $\sigma_{\rm max}$ and $k_{\rm crit}$ for different adiabatic indices are given in Table \ref{tab:2}. 

\begin{figure}[htbp] 
   \centering
   \includegraphics[width=0.32\textwidth]{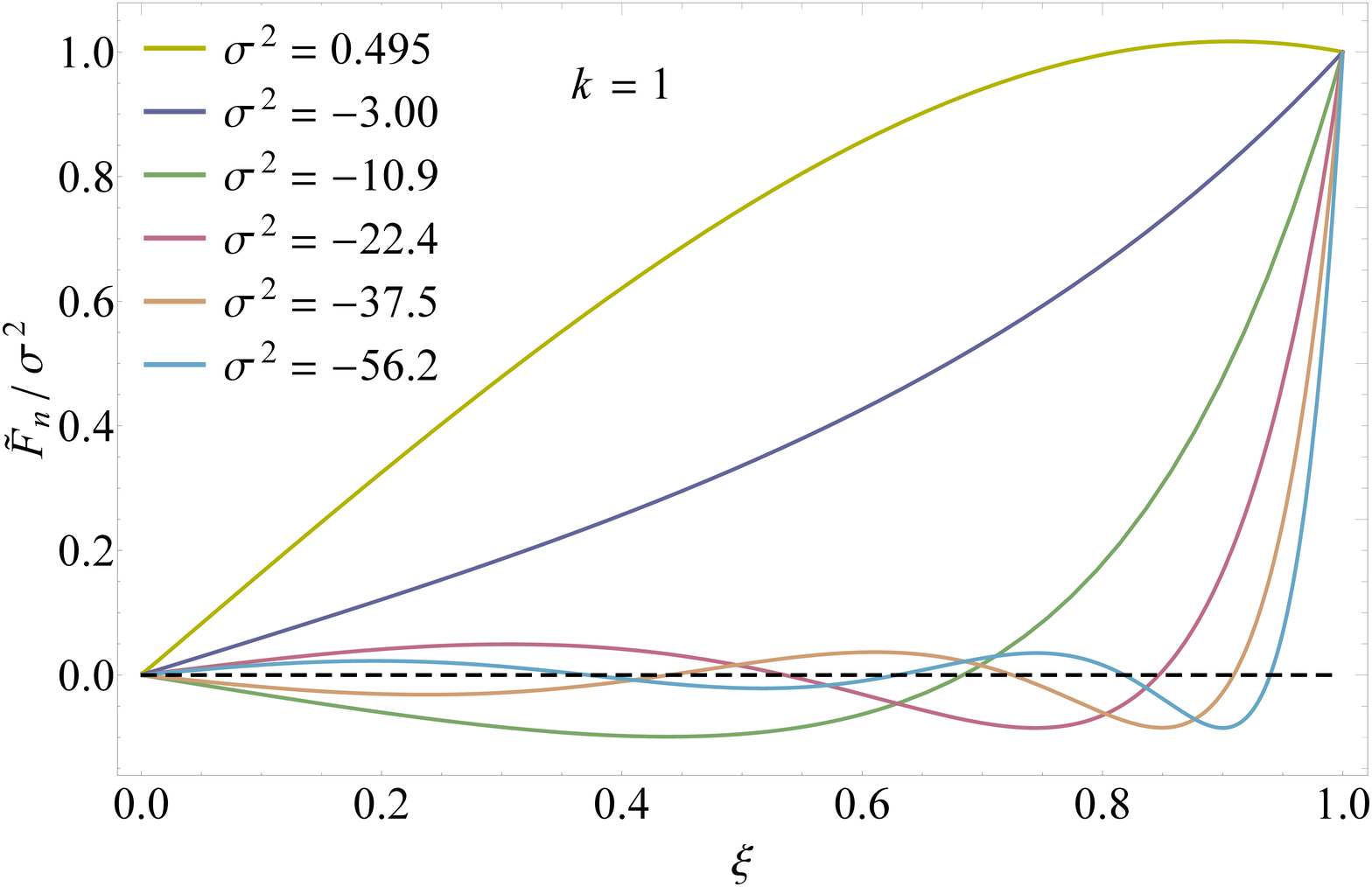} 
   \includegraphics[width=0.32\textwidth]{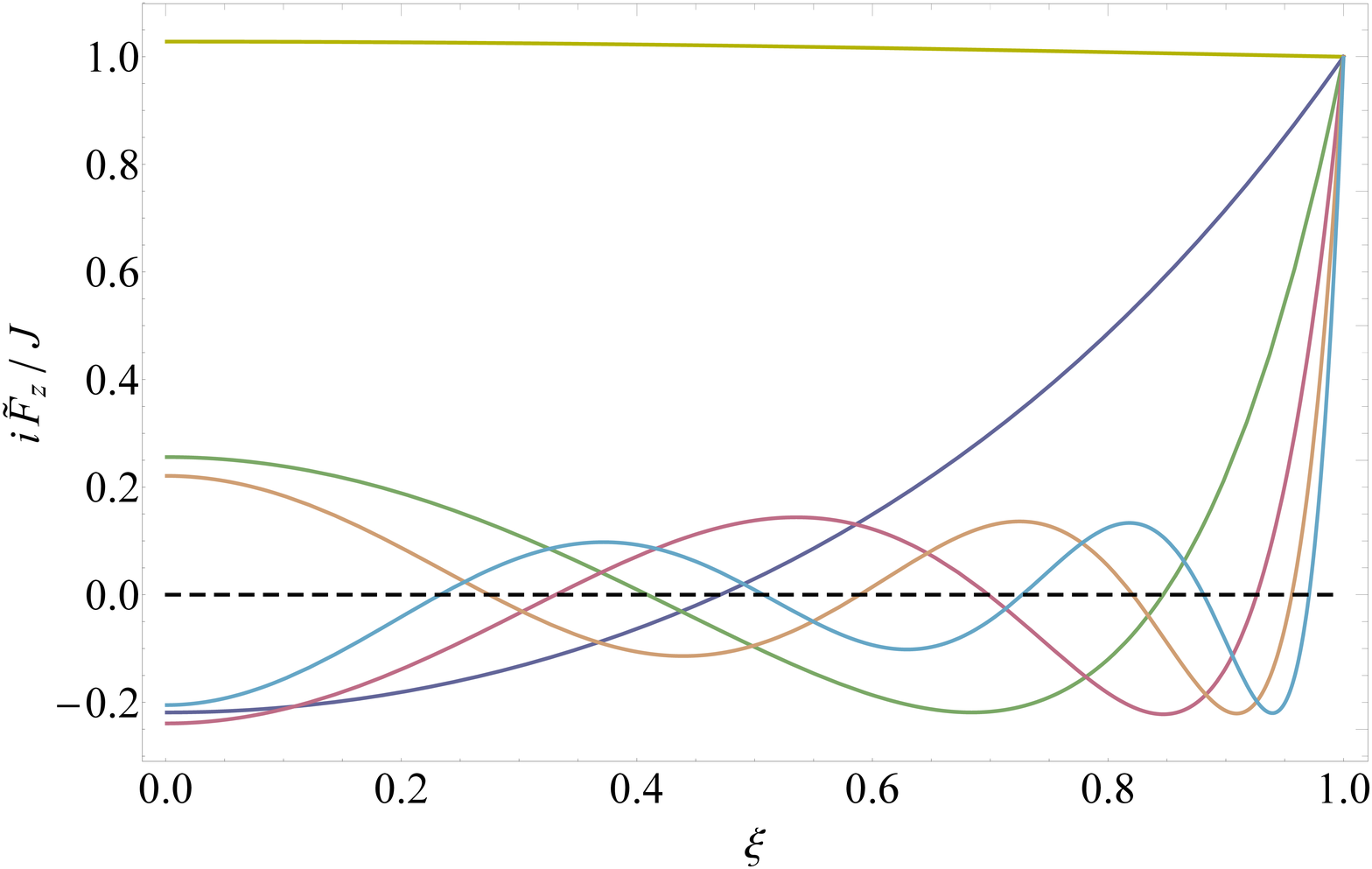} 
   \includegraphics[width=0.32\textwidth]{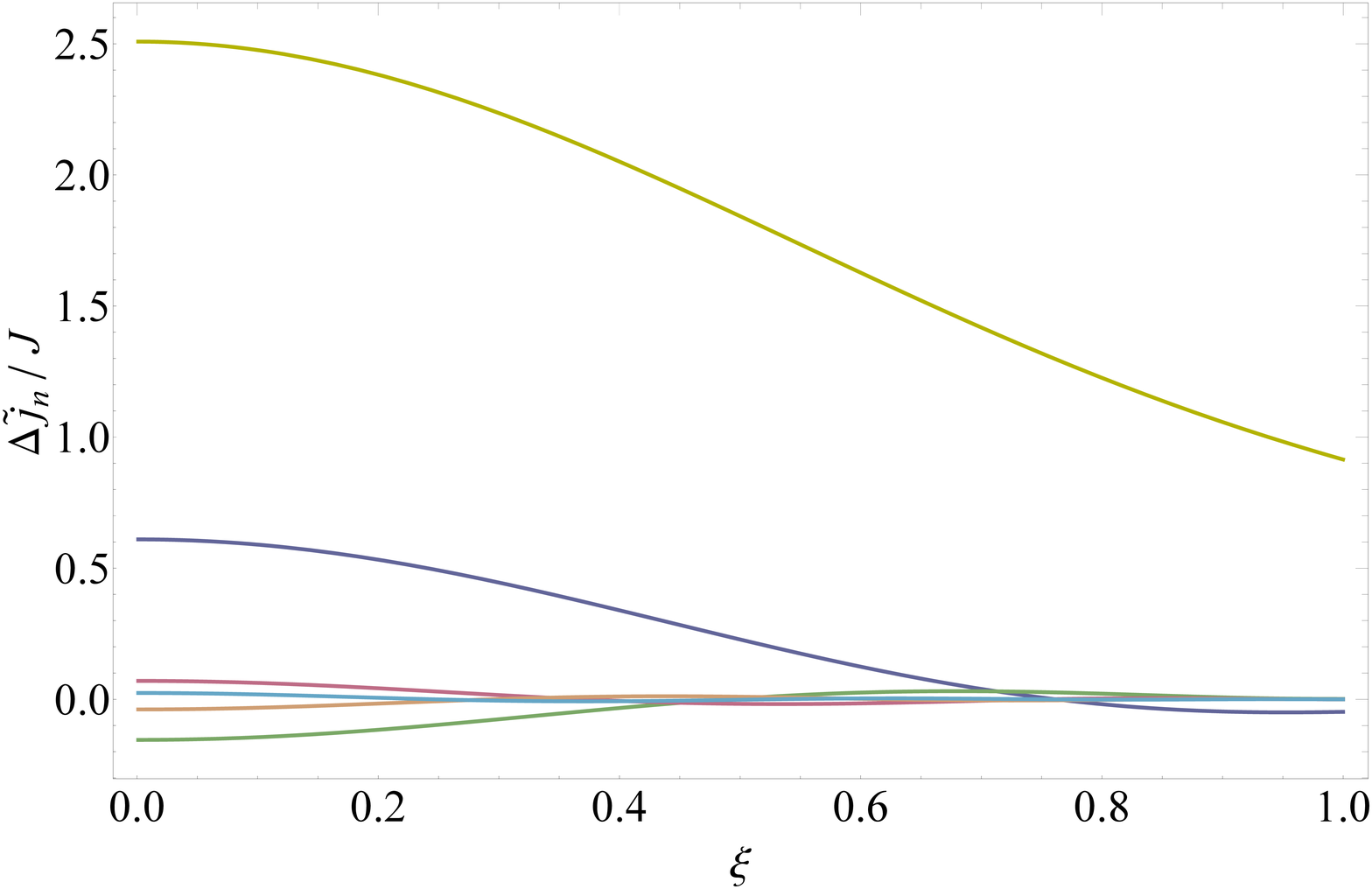} 
   \caption{The left panel shows the cylindrical-radial velocity eigenfunctions for the first seven eigenvalues shown in the legend and a wavenumber of $k = 1$, where for ease of visualization we normalized the functions by the square of the eigenvalue. The middle panel shows the eigenfunctions for the $z$-component of the velocity for the same $k$ and eigenvalues, and here we normalized the functions by the eigenvalue $J$. Comparing these two panels, we see that the unstable mode (yellow curves) plays the role of the $f$-mode (the first stable mode) when applied to the $z$-component of the velocity. The right panel gives the total change to the gravitational potential for the same eigenvalues, and demonstrates that while the first two modes possess a fairly substantial change to the gravitational field, it is effectively zero for the higher-order modes; this is a direct demonstration of the validity of Cowling's approximation.}
   \label{fig:Feigens}
\end{figure}

Figure \ref{fig:Feigens} shows the first seven eigenfunctions describing the cylindrical-radial velocity for $\gamma = 5/3$ and $k = 1$ (left panel) and the $z$-component of the velocity (middle panel). These plots illustrate the physical relevance of the additional mode $\sigma_{\rm u}$ that is unstable for this set of parameters: just as the $f$ mode (the first stable mode) describes purely outward or inward radial motion, and correspondingly the eigenfunction has no zero crossings in the range $\xi \in (0,1]$, the mode $\sigma_{\rm u}$ characterizes motion that is unidirectional along the axis of the cylinder. This feature of the eigenmodes also gives insight into the physical nature of the instability, as the root $\sigma^2 = 0$ for $k = 0$ that appears in Figure \ref{fig:kunstable} for all $\gamma$ can be understood as uniform motion along the axis of the cylinder, and therefore must exist owing to the Galilean invariance of the fluid equations. For small $k$, we can then think of the motion along the stream as comprised of two half-cylinders joined by a node in the velocity, and the fluid either converges toward or diverges from the node owing to the sinusoidal dependence of the perturbation in $z$. When the fluid converges toward (diverges from) the node, the gravitational potential increases (decreases) as the density at the node increases (decreases), and this serves to further accelerate the motion toward (away from) the node. As $k$ increases, the density is more drastically perturbed along the cylinder but the total amount of mass involved in each wavelength of the perturbation decreases, which causes the pressure to increase more drastically than, and counteract the destabilizing influence of, the gravitational potential; this in turn causes the growth rate of the instability to saturate and then decline. At the critical wavenumber $k_{\rm crit}$, the increase in the pressure exactly balances the increase in the gravitational potential, and the collapse along the axis of the cylinder proceeds at a constant velocity. This instability is therefore the cylindrical analog of the gravitational instability of a $\gamma = 4/3$ spherical polytrope, but here the instability operates along the axis of the cylinder instead of radially. 

We see from Figure \ref{fig:kp1_k1_k2} and Table \ref{tab:1} that as $\sigma^2$ become larger, the eigenvalue that constrains the gravitational potential at the surface of the cylinder, $J_{\rm n}$, becomes better approximated by $J_{\rm n} = 1$. This feature of the solutions directly validates Cowling's approximation, because to leading order in the perturbation to the surface $\tilde{\zeta}$, the gravitational potential is given by

\begin{equation}
\Phi = \sqrt{G\Lambda}\left\{j_0\left(\frac{s}{H}\right)+j_1\left(\frac{s}{H},\eta,\varphi,\tau\right)\right\} \simeq \sqrt{G\Lambda}\left\{j_0\left(\frac{s}{H_0}\right)+\Delta \tilde{j}_1\right\}, \quad \Delta \tilde{j}_1 =\tilde{j}_1 -\tilde{\zeta}\xi\frac{\partial j_0}{\partial \xi}.
\end{equation}
The last line follows from a first-order Taylor series expansion of the unperturbed potential after accounting for the corrections to the surface of the cylinder. The unperturbed potential satisfies $\partial j_0/\partial \xi(\xi = 1) = 1$, and therefore as $J_{\rm n} \rightarrow 1$, the total correction to the gravitational potential at the surface vanishes. The right panel of Figure \ref{fig:Feigens} shows the correction to the gravitational potential for the eigenmodes, $\Delta \tilde{j}_{\rm n} = \tilde{j}_{\rm n}-\xi\partial j_0/\partial \xi$, normalized by $J_{\rm n}$, and demonstrates that the total change in the gravitational potential effectively vanishes for the higher-order modes. 

\begin{table}[h]
\begin{center}
\begin{tabular}{|c|c|c|c|c|}
\hline    
\diagbox[dir =SE, outerleftsep=-.18cm, innerleftsep=2.5cm, innerwidth=.75cm, innerrightsep=.01cm, outerrightsep=-.28cm]{\hspace{-3.5cm} { Mode}}{  Wavenumber} & $k = 0.1$ & $k = 1$ & $k = 2$ & $k = 3$ \\
\hline
$u$ & $\{\sigma^2,J\} = {\{0.0267, 11.1\}}$  & $\{0.495, 11.7\}$ & \{0.160, -0.428\} & \{-1.08, 0.734\} \\ 
\hline
$f$ & \{-2.69, 0.997\} & \{-3.00, 0.954\} & \{-3.74, 0.960\} & \{-4.91, 0.977\} \\ 
\hline
$p_1$ & \{-10.7, 1.00\} & \{-10.9, 1.00\} & \{-11.6, 1.00\} & \{-12.8, 1.00\} \\ 
\hline
$p_2$ & \{-22.2, 1.00\} & \{-22.4, 1.00\} & \{-23.1, 1.00\} & \{-24.4, 1.00\} \\ 
\hline
$p_3$ & \{-37.3, 1.00\}& \{-37.5, 1.00\} & \{-38.3, 1.00\} & \{-39.5, 1.00\} \\ 
    \hline
$p_4$ & \{-56.0, 1.00\} & \{-56.2, 1.00\} & \{-56.9, 1.00\} & \{-58.1, 1.00\} \\ 
\hline
\end{tabular}
\end{center}
\caption{The eigenvalues $\sigma_{\rm n}^2$ and $J_{\rm n}$ for $\gamma = 5/3$ polytropic cylinders. The wavenumber of the perturbation is shown in the top row. }
\label{tab:1}
\end{table}

\begin{table}[h]
\begin{center}
\begin{tabular}{|c|c|c|c|c|}
\hline
$\gamma = 4/3$ & $\gamma = 1.4$ & $\gamma = 1.5$ & $\gamma = 5/3$ & $\gamma = 2$ \\
\hline
$\sigma_{\rm max} = 1.3$ & 1.1 & 0.88 & 0.72 & 0.57 \\ 
\hline
$k_{\rm max} = 2.0$ & 1.7 & 1.5 & 1.2 & 0.96 \\
\hline
 $k_{\rm crit} = 3.80$ & 3.19 & 2.65 & 2.17 & 1.75 \\ 
\hline
\end{tabular}
\end{center}
\caption{For the adiabatic indices shown in the top row, the maximum growth rate of the unstable mode ($\sigma_{\rm max}$, first row), the wavenumber at which this maximum growth rate is achieved ($k_{\rm max}$, second row), and the critical wavvenumber above which the oscillations are stable ($k_{\rm crit}$, third row). See Figure \ref{fig:kunstable} for a graphical representation of these values.}
\label{tab:2}
\end{table}

\subsection{An example} 
\label{sec:example}
To illustrate the basic features of how the instability manifests itself in the oscillations of a cylindrical filament, here we consider a specific example where we impose an initial cylindrical-radial velocity of $\delta f_{\rm s}(\xi,z) = A\xi\cos(k_{\rm p}z)$, where $A$ is an arbitrary normalization (though to be consistent with the linear analysis it should be less than one) and $k_{\rm p}$ is a specific wavenumber. Owing to the independence of the perturbation equations \eqref{Fex1} and \eqref{jex1} on the sign of $k_{\rm p}$, in this case the time-dependent solution for the perturbation in real space is also just proportional to $\cos(k_{\rm p}z)$, and we can solve Equations \eqref{Fex1} and \eqref{jex1} with $\delta \tilde{f} = \xi$ and $k = k_{\rm p}$. For definiteness here we set $k_{\rm p} = 1$, so that the cylinder has a single unstable mode. 

Writing the perturbation to the surface of the cylinder as a product of eigenvalues and a holomorphic function (the same analysis as for the purely radial perturbations that led to Equation \ref{Vsigma}) and exploiting the dependence of the equations on $\sigma^2$, we can write the solution for the perturbation to the surface of the cylinder from Equations \eqref{Fex1} and \eqref{jex1} as

\begin{equation}
\zeta = A\sum_{n}c_{\rm n}\sinh\left(\sigma_{\rm n}\tau\right), \quad c_{\rm n} = \frac{1}{\sigma_{\rm n}}\left(\frac{\partial}{\partial\sigma^2}\left[\frac{1}{\tilde{\zeta}}\right]\bigg{|}_{\sigma^2 = \sigma_{\rm n}^2}\right)^{-1} ,\label{fssol}
\end{equation}
where $\sigma_{\rm n} = \sqrt{\sigma_{\rm n}^2}$. The modes with $\sigma_{\rm n}^2 < 0$ drive stable oscillations in response to the initial perturbation, while the unstable mode with $\sigma_{\rm n}^2 > 0$ generates an exponentially growing motion of the surface of the cylinder. 

\begin{figure}[htbp] 
   \centering
   \includegraphics[width=0.495\textwidth]{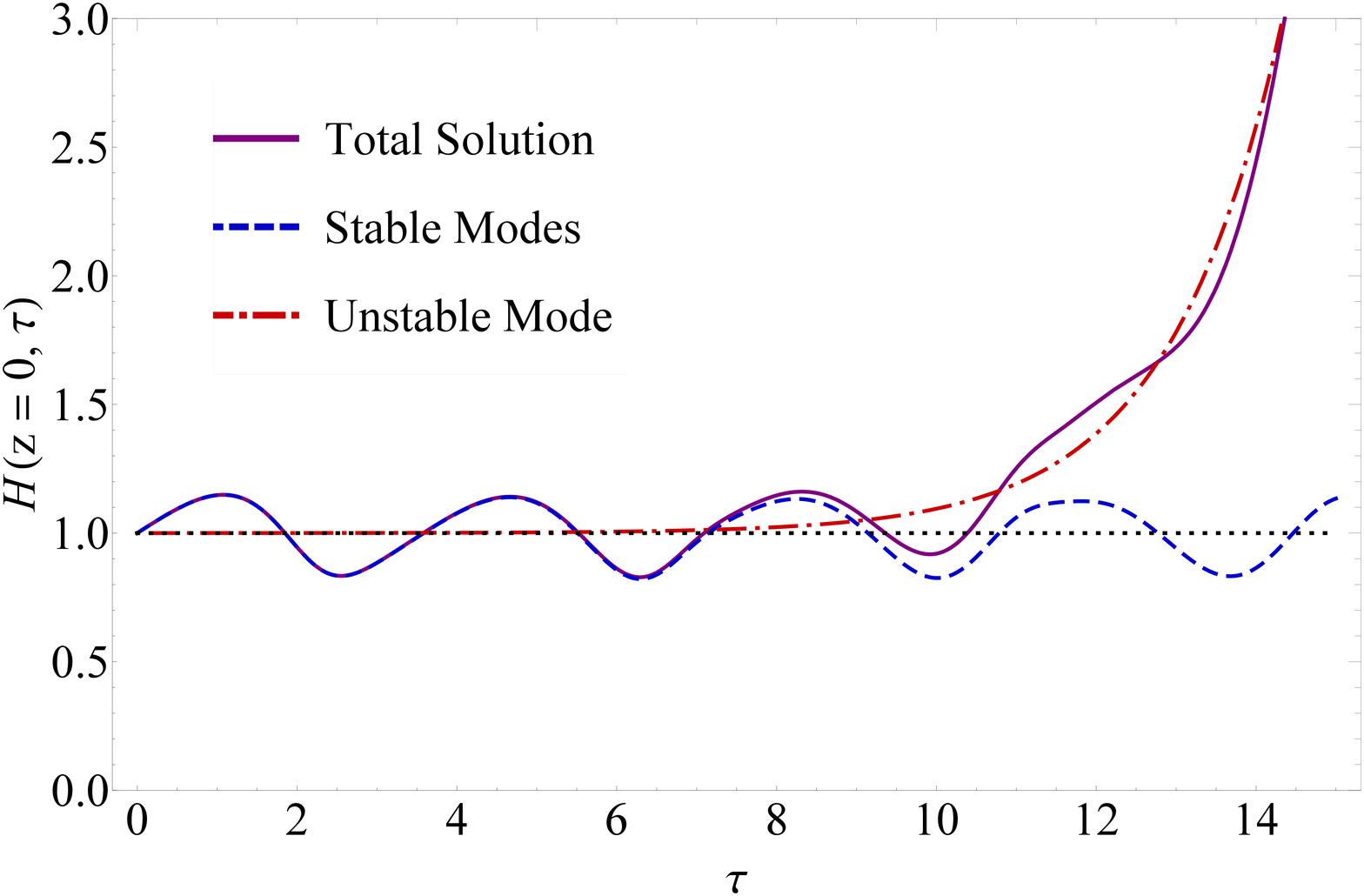} 
   \includegraphics[width=0.495\textwidth]{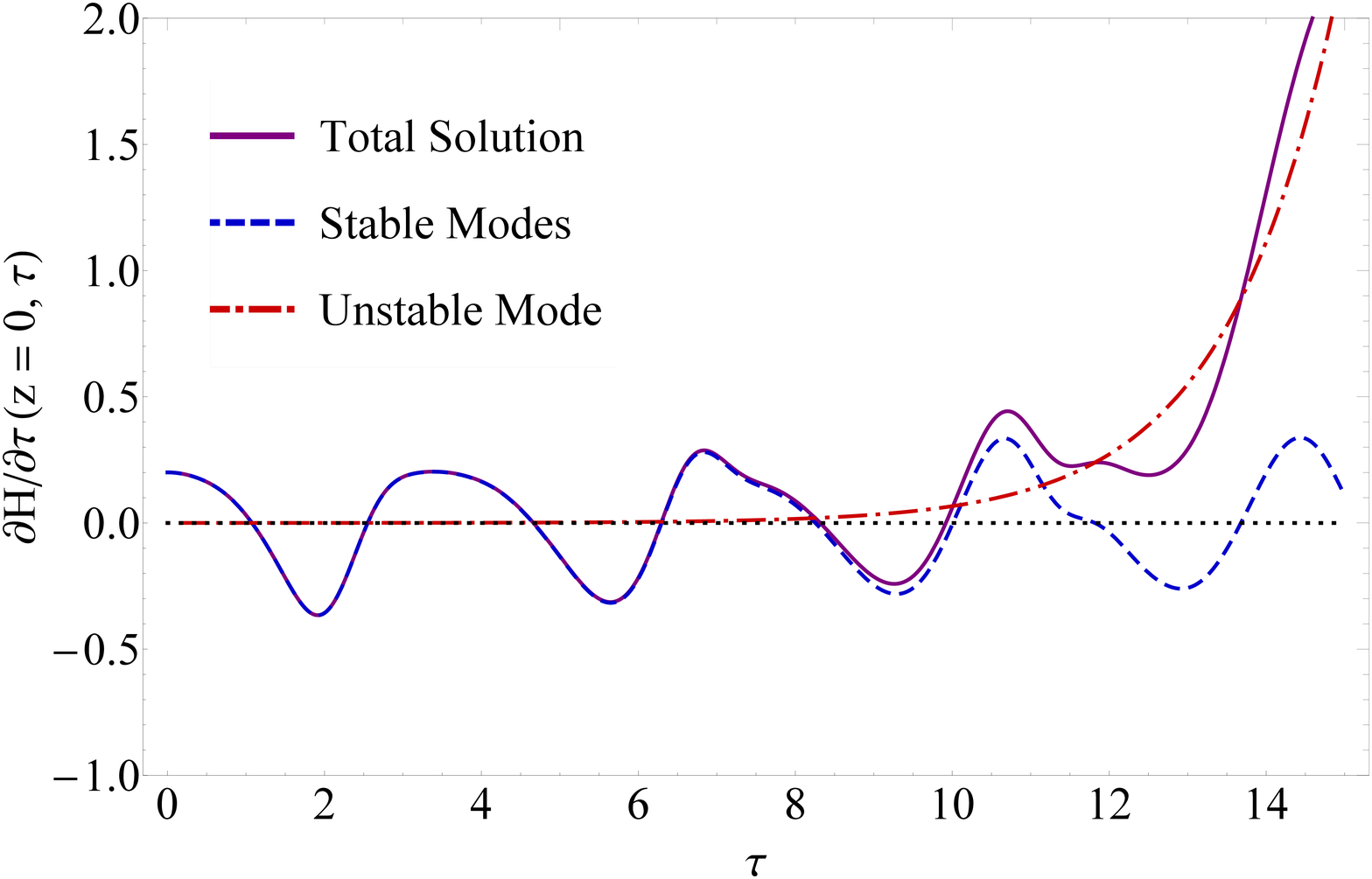} 
   \caption{The left panel shows the evolution of the displacement of the surface of the cylinder at $z = 0$ given an initial perturbation, while the right panel shows the evolution of the surface velocity, both as functions of time normalized by the sound crossing time $\tau$. The blue curves are the contribution to the solution from the stable modes, while the red curves arise from the unstable mode, and the purple curves are the total solution. Because the coefficients of the stable modes in the eigenmode decomposition are much larger than the unstable mode, the cylinder oscillates for $\sim 10$ sound crossing times before the instability sets in and runaway growth occurs. }
   \label{fig:k1perts}
\end{figure}

We can numerically determine the coefficients $c_{\rm n}$ appearing in Equation \eqref{fssol} by calculating the solution for $1/\tilde{\zeta}$ in Equations \eqref{Fex1} and \eqref{jex1} when $\sigma^2 = \sigma_{\rm n}^2+\Delta\sigma^2$, where $\Delta\sigma^2$ is a small number, and using the definition of the derivative. In practice we calculate $1/\tilde{\zeta}$ at the four points $\pm \Delta \sigma^2$, $\pm 2\Delta \sigma^2$ and use a five-point stencil to estimate the derivative; for the first three stable eigenmodes we use $\Delta \sigma^2 = 0.01$, for the fourth and fifth stable modes $\Delta \sigma^2 = 0.001$, and for the unstable mode $\Delta \sigma^2 = 0.0001$. We solve the equations with a method that is analogous to what was done to calculate the eigenmodes: for a given $\sigma$, we first determine the $1/\tilde{\zeta}$ that satisfies the boundary condition on the velocity at the origin over a range of $\tilde{J}$. We then perform the same procedure to find the $1/\tilde{\zeta}$ that satisfies the boundary condition on the potential at the origin over the same range in $\tilde{J}$. These two solutions for $1/\tilde{\zeta}$ then trace out curves as functions of $\tilde{J}$, which intersect at the unique combination of $\tilde{J}$ and $1/\tilde{\zeta}$ that satisfy the differential equations and the boundary conditions at the axis of the cylinder.

The coefficients we find for the unstable mode and the first five stable modes are, respectively, $\sigma_{\rm n}c_{\rm n} = 5.84\times10^{-4}$, 1.30, $-0.401$, 0.127, $-0.0371$, and 0.0102. The left panel of Figure \ref{fig:k1perts} shows the evolution of the surface of the cylinder $H = 1+\zeta(\tau)$ at $z = 0$ as a function of time with the amplitude of the perturbation set to $A = 0.2$, so that the initial velocity of the surface is 20\% of the sound speed; here time is in units of the sound crossing time, and the unperturbed radius of the cylinder is normalized to 1. The right panel shows the temporal evolution of the dimensionless velocity of the surface, $\partial H/\partial \tau$, at $z = 0$, and the fact that the velocity equals 0.2 at $\tau = 0$ is a consistency check on our eigenmode decomposition. The purple curves show the total solution, while the red (blue) curves give the contribution from the unstable (stable) mode(s). Because the coefficient multiplying the unstable mode is very small compared to those for the stable modes for this specific velocity perturbation, the cylinder initially appears to oscillate stably. However, after $\sim 10$ sound crossing times, the exponential growth of the unstable mode starts to dominate and the solution diverges exponentially as $\sim e^{0.7\tau}$, where the factor of $0.7$ is the unstable eigenvalue for $k = 1$ and $\gamma = 5/3$ (see Figure \ref{fig:kunstable}).  

\section{Summary and Implications}
\label{sec:summary}
In this paper we analyzed the eigenmodes of an adiabatic, polytropic cylinder that is infinite along its axis, the hydrostatic solutions for which were presented in Section \ref{sec:hse}. The eigenmodes describe the global response of such a cylinder to small perturbations, where the perturbations parameterize deviations of the fluid from the purely hydrostatic and purely cylindrical nature of the background state. When the perturbations are in the cylindrical-radial direction (Section \ref{sec:cylperts}), the modes are stable and oscillatory when the adiabatic index of the gas satisfies $\gamma > 1$. The purely radial perturbations become unstable when the cylinder is isothermal and $\gamma = 1$, but unlike the self-gravitational instability that arises when the adiabatic index of a spherical polytrope falls below $\gamma = 4/3$, the instability operates more similarly to convection (see the discussion at the end of Section \ref{sec:cylperts} and the right panel of Figure \ref{fig:eigens}).

For more general perturbations (Section \ref{sec:general}), we showed that these cylinders are characterized by a single, unstable and growing mode when the perturbations are azimuthally symmetric (no variation around the axis of the cylinder) and the wavenumber of the perturbation along the axis of the cylinder is below a critical value, $k_{\rm crit}$ (Section \ref{sec:m0modes}). Below this critical wavenumber, which is measured relative to the radius of the hydrostatic cylinder, the perturbations to the cylinder grow as $\propto e^{\sigma_{\rm u}\tau}$, where $\tau$ is time relative to the sound-crossing time over the diameter of the cylinder and $\sigma_{\rm u}$ is real and positive. We calculated the growth rate $\sigma_{\rm u}$ for the range of $k$ over which the solutions are unstable for a number of adiabatic indices, and showed that the growth rate of the instability peaks at $\sigma_{\rm max} \sim 1$ at a wavenumber $k_{\rm max} \sim 1$; see Figure \ref{fig:kunstable} and Table \ref{tab:2}. 

The eigenfunction for the unstable mode is the analog of the $f$-mode for the velocity along the axis of the cylinder, and therefore characterizes motion that is unidirectional along the cylinder; the fact that $\sigma_{\rm u}^2\rightarrow 0$ as $k \rightarrow 0$ in Figure \ref{fig:kunstable} is therefore representative of the Galilean invariance of the fluid equations and corresponds to uniform motion in one direction. We also argued that this feature of the unstable mode implies that the instability arises from the self-gravity of the gas, as long-wavelength perturbations in the velocity along the axis of the cylinder cause large portions of the fluid to converge near a node (in the velocity). This convergence of the matter drives an increase in the gravitational field, which further accelerates the gas toward the node and amplifies the motion, ultimately fueling the instability. At larger wavenumbers (smaller wavelengths) the increase in the pressure that accompanies the increase in the density at the node starts to counteract the increasing gravitational potential, and at the critical wavenumber $k_{\rm crit}$ the fluid collapses along the axis toward a node at a constant velocity. Above the critical wavenumber the pressure rises sufficiently rapidly to withstand the destabilizing influence of the self-gravity, and the cylinder stably oscillates.

Figure \ref{fig:kunstable} shows that for a specific $k < k_{\rm crit}$ and adiabatic index, the growth rate is a specific number on the order 1. Thus, if we impose a perturbation of exactly one wavenumber, e.g., a sinusoidal variation in the velocity along the cylinder, then the instability will grow at precisely the rate appropriate to that wavenumber. We calculated the evolution of the surface of an adiabatic, gas-pressure dominated filament (i.e., one with an adiabatic index of $\gamma = 5/3$) to such a sinusoidal variation, and found that the coefficient multiplying the unstable mode in the eigenmode expansion was much smaller than those multiplying the stable modes. Thus, for the specific perturbation we considered, the cylinder oscillated seemingly-stable for $\sim 10$ sound crossing times before the runaway growth ensued. In this case and after the initial oscillatory phase, the deformation of the surface grew exponentially at precisely the rate initiated by the $k = 1$ perturbation, or, from Figure \ref{fig:kunstable}, as $\propto e^{0.7\tau}$. 

However, the formation of a filament in an astrophysical context is unlikely to be accompanied by a perturbation at a single wavelength, and the Fourier decomposition of, say, the density along the filament will likely have significant power over a broad range of wavenumbers. Consequently, all of the unstable modes will start to grow at their own respective rates, augmenting the corresponding Fourier coefficients. Provided that there is not too much power at a single unstable wavelength, the most unstable mode will eventually come to dominate at late enough times and will characterize the power spectrum of the objects that gravitationally condense out of the filament. If, on the other hand, there is initially a large concentration of power at a single wavelength, then that mode may reach a nonlinear amplitude prior to the dominance of the most unstable mode, causing structure to collapse and form at that (non-maximal) wavenumber. Even in this scenario, however, the wavenumber that characterizes the distribution of collapsed objects will be less than the critical one, $k_{\rm crit}$, as above such a wavenumber the perturbations are stable. These results therefore predict that there is a preferred length scale that separates the objects that condense out of the gravitational instability of the filament and, at the very least, that there is a minimum separation between such objects.

Furthermore, the exponential growth of the instability implies that the time to reach nonlinear amplitudes is $\tau_{\rm coll} \simeq N / \sigma_{\rm max}$, where $N$ is a number that varies logarithmically with the magnitude of the initial perturbation. Because of this very weak scaling on the properties of the initial perturbation and the fact that $\sigma_{\rm max} \sim 1$, the timescale for objects to collapse out of the filament is $\tau_{\rm coll} \simeq few\times H_0/\sqrt{G\Lambda}$, where $H_0$ is the initial radius of the filament and $\Lambda$ is the initial mass per unit length. Since the wavenumber at which the growth rate is maximized is also of the order unity, we expect the mass distribution of the objects to peak at a value of $M_{\rm coll} \simeq \Lambda H_0$.

We propose that this instability is the underlying mechanism responsible for the fragmentation of debris streams formed from the tidal disruption of stars by supermassive black holes \citep{coughlin15, golightly19}. We note, however, that it is not strictly correct to apply our results (e.g., the growth timescale and mass scale) directly to tidally disrupted debris streams, as the background state of the gas comprising such a stream is not hydrostatic, but instead possesses a diverging velocity profile from the center of mass that is induced by the tidal field of the black hole. As shown by \citet{coughlin16} and \citet{coughlin16b}, a necessary condition for the stream to be gravitationally unstable in the presence of such a background shear is that the equation of state satisfy $\gamma \ge 5/3$, which arises from the competition between the shear timescale and the sound crossing time over the radius of the stream. When the equation of state satisfies this inequality, the relevant timescale that characterizes the growth of the perturbations is still given by the sound crossing time over the radius of the stream, but in this case the background line mass declines as $\Lambda \propto (t/t_{\rm dyn})^{-2/3}$ and the radius expands as $H \propto (t/t_{\rm dyn})^{2\left(\gamma-2\right)/(3\left(\gamma-1\right))}$, where $t_{\rm dyn}$ is the dynamical time at the tidal radius (being roughly the same as the sound crossing time over the initial stellar diameter; see Equations 60 and 62 of \citealt{coughlin16}). Thus, instead of growing as exponentials in time, perturbations grow exponentially to a \emph{power} of time with the power-law index less than one. Specifically, using the relation between $d\tau$ and $dt$ in Equation \eqref{dtaudt} and these scalings for $H$ and $\Lambda$, we find that 

\begin{equation}
\tau \propto \frac{1}{1-\frac{2}{3}\frac{1}{\gamma-1}}\left(\frac{t}{t_{\rm dyn}}\right)^{1-\frac{2}{3}\frac{1}{\gamma-1}}, \label{tautde}
\end{equation}
and the fastest-growing mode grows as $e^{\sigma_{\rm max}\tau}$. 

When $\gamma = 5/3$, Equation \eqref{tautde} is no longer valid, and returning to Equation \eqref{dtaudt} and letting $\gamma = 5/3$ shows that $\tau \propto \ln\left(t/t_{\rm dyn}\right)$. When the stream is gas-pressure dominated, the instability therefore only grows as a power-law in time. An additional complication in this case arises from the fact that the dynamical timescale at the center of mass ($\propto R^{3/2} \propto t$, where $R$ is the position of the marginally-bound radius) scales identically with the sound-crossing time over the stream, which implies that there is an additional parameter -- the ratio of the stream density to the black hole density at the tidal radius -- that enters into the stability analysis. In the limit that the stream density is much greater than the black hole density\footnote{It may seem as though this cannot ever be achieved, as the successful tidal disruption of the star requires that the stellar density be less than the black hole density at the tidal radius. However, it is approximately the equality between the \emph{average} stellar density by volume and the black hole density that determines the tidal radius, and hence the density at the center of mass of the stream can be substantially larger than the black hole density (e.g., see Figure 2 of \citealt{coughlin15}). The dynamical focusing of the material in the plane of the disruption can also augment the density post-pericenter \citep{coughlin16b, steinberg19}. }, the instability grows as $\propto t^{\,\sigma_{\rm max}}$, but for ratios closer to unity the eigenvalue and its dependence on the ratio must be determined from the stability analysis. It is also likely that once the ratio falls below $\sim 1$ the stream is stabilized by the tidal shear of the black hole. We defer a detailed analysis of this case to a future investigation. 

Because of this slower growth rate, it likely takes a considerable amount of time for the perturbations to become nonlinear and collapse out of the stream in a tidal disruption event, and the time taken to become nonlinear is more sensitive to the size of the initial perturbation. Consequently, even though the dynamical time in a typical tidal disruption event is on the order of hours, it may take on the order of months to years for the fragmentation to occur. We speculate that this large discrepancy in timescale between the initial stellar disruption, the formation of the clumps, and the return of the clumps to the black hole -- and the discrete feeding episodes -- could explain the late-time flaring observed in the galactic nucleus GSN 069 \citep{miniutti19} if the original outburst (observed c.~2010) was due to a tidal disruption event \citep{shu18}.

Because of the stiffness of the nuclear equation of state, the adiabatic index of the gas that comprises the tails of ejected material in the merger of two compact objects may be well represented by $\gamma \gtrsim 2$ \citep{shapiro83, wiringa88, rasio94}. Correspondingly, these tails should be susceptible to this gravitational instability and should collapse into small-scale knots, and this was found in numerical simulations by \citet{lee07} when the equation of state was as stiff as $\gamma =3$. The tails should still be unstable when $\gamma =2$, though the instability grows only as $\propto e^{t^{1/3}}$ (Equation \ref{tautde}) and perturbations will take longer to reach the nonlinear scale. This instability could therefore be responsible for late-time flaring observed in short gamma-ray bursts (e.g., \citealt{obrien06}) as condensed knots feed the remnant accretion flow at discrete times (see also \citealt{rosswog07}). 

As the equation of state softens, Figure \ref{fig:kunstable} shows that the range of unstable modes broadens such that the maximum wavenumber to which the filament is unstable, $k_{\rm crit}$, becomes larger (i.e., the filament is unstable to shorter-wavelength perturbations), and the growth rate of the instability increases. This feature arises from the fact that, as the adiabatic index approaches 1, the change in the pressure is weaker for the same change in the density, and correspondingly the pressure cannot as easily resist the destabilizing nature of self-gravity. For the cool, nearly-isothermal filaments from which stars could form, this result implies that if gas pressure is the only form of pressure support that resists gravitational collapse, then regions of length scales more dissimilar from the fastest-growing one will likely be able to reach nonlinear amplitudes and the spectrum of, e.g., the masses of protostars will likely be less well-represented by a single scale. However,  if turbulent magnetic fields are the main source of pressure, for which $\gamma = 4/3$ \citep{mckee03}, then Figure \ref{fig:kunstable} predicts that the mass spectrum and spatial distribution of stars that form along filaments encode the underlying nature of the filament itself (e.g., the line mass). In particular, if the filament has a diameter $D = 2H$, then Figure \ref{fig:kunstable} (see also Table \ref{tab:2}) predicts that star-forming cores are preferentially separated by a length $2\pi H/k_{\rm max} \simeq \pi D/2 \simeq 1.5 D$. Interestingly, this prediction is consistent with the findings of \citet{zhang20}, who investigated the properties of a set of star-forming filaments and found that the average spacing between cloud cores was $\sim 0.15$ pc, while the full-width half-max width of the filament was measured to be $0.1$ pc.

 Our analysis pertained to a polytropic cylinder in vacuum, without magnetic fields, and without rotation. The global instability we described is fundamentally due to self-gravity and the geometry of the hydrostatic configuration of the fluid, and exists in absence of these other quantities. On the other hand, if, for example, a large amount of shear is present between the filament and a background medium and the density contrast is not too low, then the Kelvin-Helmoltz instability could operate more rapidly than the instability identified here. This is precisely the context of the numerical study presented in \citet{aung19}, where the body modes of the filament operate alongside the instability driven by the shear between the surface of the cylinder and the ambient gas.

We focused exclusively on the case where the fluid is a pure polytrope, such that the adiabatic and polytropic indices are identical. If one breaks this condition, then there are more modes that characterize the oscillations of the filament, and as discussed in \citet{breysse14} these modes can drive convective instabilities when the adiabatic index is less than the polytropic index. However, owing to the physical nature of the gravitationally unstable mode described in this paper and the fact that it characterizes unidirectional motion along the filament, we find it unlikely that this mode no longer exists or becomes stable for non-polytropic filaments. This instability is therefore a generic feature of self-gravitating filaments.

\acknowledgements
ERC thanks Eliot Quataert for useful discussions. We thank the referee, Nir Mandelker, for an extremely thorough reading of the manuscript and detailed verification of our derivations, and for pointing out some appropriate references that we missed in an initial draft. ERC acknowledges support from NASA through the Hubble Fellowship Program, grant \#HST-HF2-51433.001-A awarded by the Space Telescope Science Institute, which is operated by the Association of Universities for Research in Astronomy, Incorporated, under NASA contract NAS5-26555. CJN is supported by the Science and Technology Facilities Council (grant number ST/M005917/1).

\appendix
\section{Derivation of the cylindrically symmetric perturbation equations}
\label{sec:cylderivation}
Here we derive the eigenvalue equation that describes the purely cylindrical-radial oscillations of a polytropic cylinder, Equation \eqref{finaleq1}. When the motions of the fluid are subsonic and we neglect the nonlinear terms in the velocity, the continuity, radial momentum, Poisson, and entropy equations are

\begin{equation}
\frac{\partial \rho}{\partial t}+\frac{1}{s}\frac{\partial}{\partial s}\left[\rho s v_{\rm s}\right] = 0, \quad \frac{\partial v_{\rm s}}{\partial t}+\frac{1}{\rho}\frac{\partial p}{\partial s} = -\frac{\partial \Phi}{\partial s}, \quad \frac{1}{s}\frac{\partial}{\partial s}\left[s\frac{\partial \Phi}{\partial s}\right] = 4\pi G\rho, \quad \frac{\partial K}{\partial t}+v_{\rm s}\frac{\partial K}{\partial s} = 0. \label{contappA}
\end{equation}
Here $s$ is the cylindrical radius, $\rho$ is the density, $p$ is the pressure, $v_{\rm s}$ is the cylindrical-radial velocity, $\Phi$ is the gravitational potential, and $K = \ln (p/\rho^{\gamma})$ is the specific entropy with $\gamma$ the adiabatic index of the gas. Introducing the definitions of the dimensionless density, pressure, velocity, line mass, radius, and time (see the discussions preceding Equations \ref{rhodef} and \ref{vsdef}) and adopting a polytropic relation between the pressure and the density,

\begin{equation}
\begin{split}
\rho = \frac{\Lambda}{4\pi H(t)^2}\left\{g_0(\xi)+g_1(\xi,\tau)\right\}, \quad p = \frac{G\Lambda^2}{4\pi H(t)^2}\left\{h_0(\xi)+h_1(\xi,\tau)\right\}, \quad v_{\rm s} = \sqrt{G\Lambda}f_{\rm s}(\xi,\tau), \\
 \lambda_0(\xi) = \int_0^{\xi}g_0(\xi)\,\xi\,d\xi, \quad \lambda_1(\xi,\tau) = \int_0^{\xi}g_1(\xi,\tau)\,\xi\,d\xi, \quad \xi = \frac{s}{H(t)}, \quad d\tau = \frac{\sqrt{G\Lambda}}{H(t)}dt, \quad h_0 = K_0 g_0^{\gamma}, \label{gendefs}
 \end{split}
\end{equation} 
and inserting these into Equations \eqref{contappA} yield the equation of hydrostatic balance for the unperturbed quantities,

\begin{equation}
\frac{1}{g_0}\frac{\partial h_0}{\partial \xi} = -\frac{\lambda_0}{\xi}, \label{lane2}
\end{equation}
and the following three equations for the perturbed quantities:

\begin{equation}
\frac{\partial \lambda_1}{\partial \tau}+f_{\rm s}\frac{\partial \lambda_0}{\partial \xi} = V(\tau)\xi\frac{\partial \lambda_0}{\partial \xi}, \quad \frac{\partial f_{\rm s}}{\partial \tau}-\frac{g_1}{g_0^2}\frac{\partial h_0}{\partial \xi}+\frac{1}{g_0}\frac{\partial h_1}{\partial \xi} = -\frac{\lambda_1}{\xi}, \quad \frac{\partial}{\partial \tau}\left[\frac{h_1}{h_0}-\frac{\gamma g_1}{g_0}\right] = -2\left(\gamma-1\right)V(\tau). \label{lambda1}
\end{equation}
Here

\begin{equation}
V = \frac{1}{\sqrt{G\Lambda}}\frac{\partial H}{\partial t}
\end{equation}
is the dimensionless surface velocity, and using the polytropic relation between the pressure and the density and the definition of the line mass yields the Lane-Emden equation from Equation \eqref{lane2} (cf.~Equation \ref{hse}). We also used the following relations that result from the transformations between partial derivatives with respect to physical coordinates and dimensionless coordinates:

\begin{equation}
\frac{\partial}{\partial t} = \frac{\partial\tau}{\partial t}\frac{\partial}{\partial \tau}+\frac{\partial \xi}{\partial t}\frac{\partial}{\partial \xi} = \frac{\sqrt{G\Lambda}}{H}\frac{\partial}{\partial \tau}-\frac{1}{H}\frac{\partial H}{\partial t}\xi\frac{\partial}{\partial \xi}, \quad \frac{\partial}{\partial s} = \frac{\partial \xi}{\partial s}\frac{\partial}{\partial \xi} = \frac{1}{H}\frac{\partial}{\partial \xi}.
\end{equation}
Taking the Laplace transform of the first and third of Equations \eqref{lambda1}, where the Laplace transform of $\lambda_1$ is

\begin{equation}
\tilde{\lambda}_1(\xi,\sigma) = \int_0^{\infty}\lambda_1(\xi,\tau)e^{-\sigma\tau}d\tau,
\end{equation}
and similarly for other quantities, and assuming for simplicity that there is no initial perturbation to the line mass or the entropy, we can solve for the Laplace-transformed perturbation to the line mass and the density:

\begin{equation}
\tilde{\lambda}_1 = \frac{1}{\sigma}\left(\tilde{V}\xi-\tilde{f}_{\rm s}\right)\xi g_0, \quad \tilde{h}_1 = \frac{\gamma h_0}{g_0}\tilde{g}_1-2\left(\gamma-1\right) h_0\frac{1}{\sigma}\tilde{V}. \label{lambda1eq}
\end{equation}
Taking the Laplace transform of the second of Equation \eqref{lambda1}, letting there be an initial velocity perturbation $\delta f(\xi)$, and using these solutions for $\tilde{\lambda}_1$ and $\tilde{h}_1$ gives

\begin{equation}
\sigma \tilde{f}_{\rm s}-\frac{\tilde{g}_1}{g_0^2}\frac{\partial h_0}{\partial \xi}+\frac{1}{g_0}\frac{\partial}{\partial \xi}\left[\frac{\gamma h_0}{g_0}\tilde{g}_1-2\left(\gamma-1\right)h_0\frac{\tilde{V}}{\sigma}\right] = -\frac{1}{\sigma}\left(\tilde{V}\xi-\tilde{f}_{\rm s}\right)g_0+\delta f.
\end{equation}
Using the fact that $\xi\tilde{g}_1 = \partial\tilde{\lambda}_1/\partial\xi$ from Equation \eqref{gendefs}, using Equation \eqref{lambda1eq} to remove the dependence on $\tilde{\lambda}_1$, and a few additional algebraic manipulations turns this into Equation \eqref{finaleq1}, being the second order differential equation for $\tilde{f}_{\rm s}$. 

\section{Derivation of the generic eigenvalue equations}
\label{sec:derivation}
Here we derive the eigenvalue equations that describe general (i.e., not restricted to the cylindrical-radial direction $s$) oscillations of an adiabatic cylinder, specifically Equations \eqref{Fex1} and \eqref{jex1}. In the perturbative limit where we neglect the nonlinear terms in the velocity and gradients in the $z$ and $\varphi$ directions, where $z$ is along the axis of the cylinder and $\varphi$ is the azimuthal angle around the axis, the continuity, momentum, entropy, and Poisson equations are

\begin{equation}
\begin{split}
\frac{\partial \rho}{\partial t}+\frac{1}{s}\frac{\partial}{\partial s}\left[s\rho v_{\rm s}\right]+\frac{\partial}{\partial z}\left[\rho v_{\rm z}\right]+\frac{1}{s}\frac{\partial}{\partial \varphi}\left[\rho v_{\varphi}\right] = 0, \quad \frac{\partial v_{\rm s}}{\partial t}+\frac{1}{\rho}\frac{\partial p}{\partial s} = -\frac{\partial \Phi}{\partial s}, \quad \frac{\partial v_{\rm z}}{\partial t}+\frac{1}{\rho}\frac{\partial p}{\partial z} = -\frac{\partial \Phi}{\partial z}, \\
\frac{\partial v_{\varphi}}{\partial t}+\frac{1}{s}\frac{1}{\rho}\frac{\partial p}{\partial \varphi} = -\frac{1}{s}\frac{\partial \Phi}{\partial \varphi}, \quad \frac{\partial K}{\partial t}+v_{\rm s}\frac{\partial K}{\partial s} = 0, \quad \frac{1}{s}\frac{\partial}{\partial s}\left[s\frac{\partial \Phi}{\partial s}\right]+\frac{\partial^2\Phi}{\partial z^2}+\frac{1}{s^2}\frac{\partial ^2\Phi}{\partial \varphi^2} = 4\pi G\rho
\end{split}
\end{equation}
Here $\rho$ is the fluid density, $p$ is the pressure, $v_{\rm s}$, $v_{\rm z}$, and $v_{\varphi}$ are the $s$, $z$, and $\varphi$ components of the velocity, $K = \ln(p/\rho^{\gamma})$ is the specific entropy with $\gamma$ the adiabatic index, and $\Phi$ is the gravitational potential. We now non-dimensionalize these equations by defining

\begin{equation}
\begin{split}
\xi = \frac{s}{H(t,z,\varphi)}, \,\,\, \eta = \frac{z}{H}, \,\,\, d\tau = \frac{\sqrt{G\Lambda}}{H}dt, \,\,\, \rho = \frac{\Lambda}{4\pi H^2}\left\{g_0(\xi)+g_1(\xi,\eta,\varphi,\tau)\right\}, \,\,\, p = \frac{G\Lambda^2}{4\pi H^2}\left\{h_0(\xi)+h_1(\xi,\eta,\varphi,\tau)\right\}, \\ \{v_{\rm s}, \, v_{\rm z}, \, v_{\varphi}\} = \sqrt{G\Lambda}\left\{f_{\rm s}(\xi,\eta,\varphi,\tau), \, f_{\rm z}(\xi,\eta,\varphi,\tau), \, f_{\varphi}(\xi,\eta,\varphi,\tau)\right\}, \quad \Phi = G\Lambda\left\{j_0(\xi)+j_1(\xi,\eta,\varphi,\tau)\right\},
\end{split}
\end{equation}
where $H(z,\varphi,t)$ is the surface of the cylinder that includes the perturbations. We further parameterize the surface of the cylinder by

\begin{equation}
H = H_0\left\{1+\zeta(\eta,\varphi,\tau)\right\}
\end{equation}
where $\zeta$ is the dimensionless perturbation to the cylinder and is assumed to be much less than one. In terms of these variables and letting the unperturbed cylinder be a polytrope, so $h_0 = K_0 g_0^{\gamma}$, the subscript-zero quantities satisfy the Lane-Emden equation \eqref{hse}, and the leading-order, linearized fluid equations become

\begin{equation}
\frac{\partial g_1}{\partial \tau}+\frac{\partial}{\partial \eta}\left[g_0f_{\rm z}\right]+\frac{1}{\xi}\frac{\partial}{\partial \varphi}\left[g_0 f_{\varphi}\right] = \frac{1}{\xi}\frac{\partial}{\partial \xi}\left[g_0\xi\left(\frac{\partial \zeta}{\partial \tau}\xi-f_{\rm s}\right)\right], \label{contapp2}
\end{equation}
\begin{equation}
\frac{\partial f_{\rm s}}{\partial \tau}-\frac{g_1}{g_0^2}\frac{\partial h_0}{\partial \xi}+\frac{1}{g_0}\frac{\partial h_1}{\partial \xi} = -\frac{\partial j_1}{\partial \xi}, \label{smomapp2}
\end{equation}
\begin{equation}
\frac{\partial f_{\rm z}}{\partial \tau}+\frac{1}{g_0}\left(-2\frac{\partial \zeta}{\partial \eta}h_0+\frac{\partial h_1}{\partial \eta}\right) = -\frac{\partial j_1}{\partial \eta}, \label{zmomapp2}
\end{equation}
\begin{equation}
\frac{\partial f_{\varphi}}{\partial \tau}+\frac{1}{g_0\xi}\left(-2\frac{\partial \zeta}{\partial \varphi}h_0+\frac{\partial h_1}{\partial \varphi}\right) = -\frac{1}{\xi}\frac{\partial j_1}{\partial \varphi}, \label{phimomapp2}
\end{equation}
\begin{equation}
\frac{\partial}{\partial \tau}\left[\frac{h_1}{h_0}-\frac{\gamma g_1}{g_0}\right] = -2\left(\gamma-1\right)\frac{\partial \zeta}{\partial \tau}, \label{entapp2}
\end{equation}
\begin{equation}\frac{1}{\xi}\frac{\partial}{\partial \xi}\left[\xi\frac{\partial j_1}{\partial \xi}\right]+\frac{\partial^2j_1}{\partial \eta^2}+\frac{1}{\xi^2}\frac{\partial^2j_1}{\partial \varphi^2}-\left(\frac{\partial^2\zeta}{\partial \eta^2}+\frac{1}{\xi^2}\frac{\partial^2\zeta}{\partial \varphi^2}\right)\xi\frac{\partial j_0}{\partial \xi} = g_1\quad \label{poissonapp2}
\end{equation}
Note that, in deriving these equations, we used the following transformations between partial derivatives:

\begin{equation}
\frac{\partial}{\partial t} = \frac{\partial \tau}{\partial t}\frac{\partial}{\partial \tau}+\frac{\partial \xi}{\partial t}\frac{\partial}{\partial \xi} = \frac{\sqrt{G\Lambda}}{H}\frac{\partial}{\partial \tau}, \,\,\, \frac{\partial}{\partial s} = \frac{\partial \xi}{\partial s}\frac{\partial}{\partial \xi} = \frac{1}{H}\frac{\partial}{\partial \xi}, \,\,\, \frac{\partial}{\partial z} = \frac{1}{H}\frac{\partial}{\partial \eta}-\frac{1}{H}\frac{\partial H}{\partial z}\xi\frac{\partial}{\partial \xi}.
\end{equation}
We further emphasize that these transformations are correct to \emph{leading, nonlinear order} in the perturbed quantities. For example, we do not have to account for the additional term $\partial\eta/\partial t\times\partial/\partial \eta$ that arises in transforming from $t$ to the dimensionless variables because this term will always be of second order in the perturbations. An analogous nonlinear term is dropped from the transformation between the derivative with respect to $z$ and the dimensionless variables. Similarly, in the Poisson equation, the only first order term that survives the second order differentiation of the unperturbed potential $j_0$ with respect to $z$ is

\begin{equation}
\frac{\partial j_0}{\partial z} = -\frac{1}{H}\frac{\partial H}{\partial z}\xi\frac{\partial j_0}{\partial \xi} \quad \Rightarrow \quad \frac{\partial ^2j_0}{\partial z^2} = -\frac{1}{H}\frac{\partial^2H}{\partial z^2}\xi\frac{\partial j_0}{\partial \xi}.
\end{equation}
It is because of this term (and the analogous one that arises from the differentiation with respect to $\varphi$) that the dependence on $\zeta$ appears in Equation \eqref{poissonapp2}.

We now take the Fourier and Laplace transform of this set of equations, where the combined transform of $f_{\rm s}$ is

\begin{equation}
\tilde{f}_{\rm s}(\xi,k,m,\sigma) = \int_0^{\infty}\int_{-\infty}^{\infty}\int_0^{2\pi}f_{\rm s}\left(\xi,\eta,\varphi,\tau\right)e^{-im\varphi-ik\eta-\sigma\tau}d\varphi\,d\eta\,d\tau.
\end{equation}
Here $k$ is a continuous variable while $m$ is restricted to positive and negative integers. We assume for simplicity that there is only an initial perturbation to the $s$-component of the velocity, the Fourier transform of which we denote $\delta\tilde{f}_{\rm s}$. Equations \eqref{zmomapp2} and \eqref{phimomapp2} can then be combined to show that

\begin{equation}
\tilde{f}_{\varphi} = \frac{m}{k}\frac{\tilde{f}_{\rm z}}{\xi}, \label{fphiint}
\end{equation}
while Equation \eqref{entapp2} can be rearranged to give

\begin{equation}
\tilde{h}_1 = \frac{\gamma h_0}{g_0}\tilde{g}_1-2\left(\gamma-1\right)h_0\tilde{\zeta}, \label{h1int}
\end{equation}
and using this relation in Equation \eqref{zmomapp2} and rearranging yields

\begin{equation}
\tilde{f}_{\rm z} = -\frac{ik}{\sigma}\left(\tilde{j}_1+\frac{\gamma h_0}{g_0}\frac{\tilde{g}_1}{g_0}-\frac{2\gamma h_0}{g_0}\tilde{\zeta}\right). \label{fzint}
\end{equation}
Using Equations \eqref{fphiint} and \eqref{fzint} in Equation \eqref{contapp2} and performing a few algebraic manipulations then shows

\begin{equation}
\frac{\tilde{g}_1}{g_0} = \frac{\frac{1}{g_0\xi}\frac{\partial}{\partial \xi}\left[g_0\xi\left(\sigma^2\tilde{\zeta}\xi-\sigma\tilde{f}_{\rm s}\right)\right]-\left(k^2+\frac{m^2}{\xi^2}\right)\left(\tilde{j}_1-\frac{2\gamma h_0}{g_0}\tilde{\zeta}\right)}{\sigma^2+\left(k^2+\frac{m^2}{\xi^2}\right)\frac{\gamma h_0}{g_0}}. \label{g1ong0}
\end{equation}
Using Equations \eqref{h1int} in Equation \eqref{smomapp2} gives

\begin{equation}
\sigma\tilde{f}_{\rm s}+\frac{\partial}{\partial \xi}\left[\frac{\gamma h_0}{g_0}\frac{\tilde{g}_1}{g_0}+\tilde{j}_1-\frac{2\gamma h_0}{g_0}\tilde{\zeta}\right] = \delta\tilde{f},
\end{equation}
and inserting Equation \eqref{g1ong0} into this relation, performing a few more algebraic manipulations and defining $\tilde{F}_{\rm s} \equiv \sigma \tilde{f}_{\rm s}$ yields Equation \eqref{Fex1}. Taking the Fourier transform of the Poisson equation \eqref{poissonapp2}, using Equation \eqref{g1ong0} and performing some rearrangements gives Equation \eqref{jex1}. Using Equation \eqref{g1ong0} in Equation \eqref{fzint} and defining $\tilde{F}_{\rm z} \equiv \sigma \tilde{f}_{\rm z}$ also shows

\begin{equation}
\tilde{F}_{\rm z} = -ik\left(\frac{\frac{1}{g_0\xi}\frac{\partial}{\partial \xi}\left[g_0\xi\left(\sigma^2\tilde{\zeta}\xi-\tilde{F}_{\rm s}\right)\right]\frac{\gamma h_0}{g_0}+\sigma^2\left(\tilde{j}_1-\frac{2\gamma h_0}{g_0}\tilde{\zeta}\right)}{\sigma^2+\left(k^2+\frac{m^2}{\xi^2}\right)\frac{\gamma h_0}{g_0}}\right), \label{eigenfz}
\end{equation}
which -- for the eigenvalues $\sigma_{\rm n}$ and eigenfunctions $\tilde{F}_{\rm n}$ and $\tilde{j}_{\rm n}$ -- are the eigenfunctions of the $z$-component of the velocity.

\section{Eulerian formulation of spherical eigenmodes}
\label{sec:spherical}
Because we have not encountered them elsewhere in the literature, they provide a check on our results, and they may also be useful to other readers in other contexts, here we provide the completely Eulerian eigenvalue equations that govern the oscillations of spherical polytropes. The fluid equations in spherical coordinates in the limit that we neglect nonlinear terms in the fluid velocity are

\begin{equation}
\begin{split}
\frac{\partial\rho}{\partial t}+\frac{1}{r^2}\frac{\partial}{\partial r}\left[r^2\rho v_{\rm r}\right]+\frac{1}{r\sin\theta}\frac{\partial}{\partial \theta}\left[\sin\theta \rho v_{\theta}\right]+\frac{1}{r\sin\theta}\frac{\partial}{\partial \phi}\left[\rho v_{\phi}\right], \,\,\, \frac{\partial K}{\partial t}+v_{\rm r}\frac{\partial K}{\partial r} = 0, \,\,\, \frac{\partial v_{\rm r}}{\partial t}+\frac{1}{\rho}\frac{\partial p}{\partial r} = -\frac{\partial \Phi}{\partial r},  \\ 
\frac{\partial v_{\theta}}{\partial t}+\frac{1}{r}\frac{1}{\rho}\frac{\partial p}{\partial \theta} = -\frac{1}{r}\frac{\partial \Phi}{\partial \theta}, \,\,\, \frac{\partial v_{\phi}}{\partial t}+\frac{1}{r\sin\theta}\frac{1}{\rho}\frac{\partial p}{\partial \phi} = -\frac{1}{r\sin\theta}\frac{\partial \Phi}{\partial \phi}, \\ 
\frac{1}{r^2}\frac{\partial}{\partial r}\left[r^2\frac{\partial \Phi}{\partial r}\right]+\frac{1}{r^2\sin\theta}\frac{\partial}{\partial \theta}\left[\sin\theta\frac{\partial \Phi}{\partial \theta}\right]+\frac{1}{r^2\sin^2\theta}\frac{\partial^2\Phi}{\partial\phi^2} = 4\pi G\rho.
\end{split}
\end{equation}
Here the definitions of the fluid variables are precisely the same as or analogous to the cylindrical case, i.e., $\rho$ is the fluid density, $p$ is the pressure, $r$ is the spherical-radial coordinate, $\theta$ is the spherical polar angle, $\phi$ is the spherical azimuthal angle, $v_{\rm r}$, $v_{\theta}$, and $v_{\phi}$ are the radial, poloidal, and azimuthal components of the velocity, $K = \ln(p/\rho^{\gamma})$ is the specific entropy with $\gamma$ the adiabatic index, and $\Phi$ is the gravitational potential. We now non-dimensionalize the fluid quantities by introducing the following definitions:

\begin{equation}
\rho(\xi,\theta,\phi,\tau) = \frac{M}{4\pi R(\theta,\phi,t)^3}\left\{g_0(\xi)+g_1(\xi,\theta,\phi,\tau)\right\}, \,\,\, p(\xi,\theta,\phi,\tau) = \frac{GM^2}{4\pi R^4}\left\{h_0(\xi)+h_1(\xi,\theta,\phi,\tau)\right\},
\end{equation}
\begin{equation}
 v_{\rm r,\theta,\phi} = \sqrt{\frac{GM}{R}}f_{\rm r,\theta,\phi}(\xi,\theta,\phi,\tau), \,\,\, \Phi(\xi,\theta,\phi,\tau) = \frac{GM}{R}\left\{j_0(\xi)+j_1(\xi,\theta,\phi,\tau)\right\}, \,\,\, \xi = \frac{r}{R}, \,\,\, d\tau = \frac{\sqrt{GM}}{R^{3/2}}dt, 
\end{equation}
Here $M$ is the total mass of the polytrope and $R(\theta,\phi,t)$ is the radius of the star at which the density equals zero, which -- including the perturbations induced from aspherical motions -- depends both on time and the polar angles. Consistent with the subsonic nature of the perturbations and the small angular variation of the surface with respect to radius, we further parameterize this radius by

\begin{equation}
R = R_0\left\{1+\zeta\left(\tau,\theta,\phi\right)\right\},
\end{equation}
where $R_0$ is the unperturbed radius of the star and $\zeta$ is a small correction induced by the perturbations. Inserting the above into the fluid equations, the zeroth-order terms can be combined to give

\begin{equation}
\frac{K_0\gamma}{\gamma-1}\frac{\partial}{\partial \xi}\left[\left(\frac{1}{\xi^2}\frac{\partial m_0}{\partial \xi}\right)^{\gamma-1}\right] = -\frac{m_0}{\xi^2}, \label{lanesph}
\end{equation}
where

\begin{equation}
m_0(\xi) = \int_0^{\xi}\tilde{\xi}^2g_0(\tilde{\xi})\,d\tilde{\xi}
\end{equation}
is the dimensionless mass enclosed within the polytrope. Equation \eqref{lanesph} is the familiar Lane-Emden equation in spherical coordinates written in terms of the dimensionless mass, and aside from geometrical factors is the same as Equation \eqref{hse}. It can be solved in precisely the same way that we solved Equation \eqref{hse} to obtain solutions for the unperturbed density, pressure and the entropy $K_0$ for different adiabatic indices $\gamma$. 

Owing to the spherical nature of the background state, we expand the angular part of the perturbation to the surface and the fluid variables in spherical harmonics; in particular, we write

\begin{equation}
\begin{split}
&g_1(\xi,\theta,\phi,\tau) = g_1(\xi,\tau)Y^{\rm m}_{\ell}(\theta,\phi), \,\,\, h_1(\xi,\theta,\phi,\tau) = h_1(\xi,\tau)Y^{\rm m}_{\ell}(\theta,\phi), \,\,\, f_{\rm r}(\xi,\theta,\phi,\tau) = f_{\rm r}(\xi,\tau)Y^{\rm m}_{\ell}(\theta,\phi), \\
& j_1(\xi,\theta,\phi,\tau) = j_1(\xi,\tau)Y^{\rm m}_{\ell}(\theta,\phi), \,\,\, f_{\theta}(\xi,\theta,\phi,\tau) = f_{\perp}(\xi,\tau)\frac{\partial Y^{\rm m}_{\ell}}{\partial \theta}, \,\,\, f_{\phi}(\xi,\theta,\phi,\tau) = f_{\perp}(\xi,\tau)\frac{1}{\sin\theta}\frac{\partial Y^{\rm m}_{\ell}}{\partial \phi}, \\
&\zeta(\theta,\phi,\tau) = \zeta(\tau)Y^{\rm m}_{\ell}(\theta,\phi).
\end{split}
\end{equation}
To avoid introducing yet more notation, we used the same symbol to represent the angle-independent functions that represent the fluid quantities, except for the $\theta$ and $\phi$-components of the velocity where we introduced the function $f_{\perp}$. Substituting the above definitions into the equations and dropping nonlinear terms leads to a self-consistent set of relations for the $\theta$-independent quantities, i.e., the $\theta$ and $\phi$ components of the momentum equation yield the same equation for $f_{\perp}$, and the fundamental equation for the spherical harmonics,

\begin{equation}
\frac{1}{\sin\theta}\frac{\partial}{\partial \theta}\left[\sin\theta\frac{\partial Y^{\rm m}_{\ell}}{\partial \theta}\right]+\frac{1}{\sin^2\theta}\frac{\partial^2 Y^{\rm m}_{\ell}}{\partial \phi^2} = -\ell\left(\ell+1\right)Y^{\rm m}_{\ell},
\end{equation}
completely removes the angular dependence in the equations. We can now take the Laplace transform of the equations and follow the same set of procedures that we did in Appendix \ref{sec:derivation} to reduce the entire set of relations to two, second-order, coupled ODEs for the perturbation to the gravitational potential and the radial component of the velocity. Defining $\sigma \tilde{f}_{\rm r} \equiv \tilde{F}_{\rm r}$ and focusing on the eigenmodes, for which $\tilde{\zeta} \propto \left(\sigma-\sigma_{\rm n}\right)^{-1}$, and letting $\tilde{F}_{\rm n} = \tilde{F}_{\rm r}/\tilde{\zeta}$, $\tilde{j}_{\rm n} = \tilde{j}_1/\tilde{\zeta}$, and $\sigma \rightarrow \sigma_{\rm n}$, the eigenmode equations are

\begin{equation}
\tilde{F}_{\rm n}+\frac{\partial}{\partial \xi}\left[\frac{\frac{1}{g_0\xi^2}\frac{\partial}{\partial \xi}\left[g_0\xi^2\left(\sigma_{\rm n}^2\xi-\tilde{F}_{\rm n}\right)\right]\frac{\gamma h_0}{g_0}+\sigma_{\rm n}^2\left(\tilde{j}_{\rm n}-\frac{3\left(\gamma-4/3\right)}{\gamma-1}\frac{\gamma h_0}{g_0}\right)}{\sigma_{\rm n}^2+\frac{\ell\left(\ell+1\right)}{\xi^2}\frac{\gamma h_0}{g_0}}\right] = 0, \label{Fnspherical}
\end{equation}
\begin{equation}
\frac{1}{\xi^2}\frac{\partial}{\partial \xi}\left[\xi^2\frac{\partial \tilde{j}_{\rm n}}{\partial \xi}\right]-\frac{\ell\left(\ell+1\right)}{\xi^2}\left(\tilde{j}_{\rm n}-\frac{\partial}{\partial\xi}\left[\xi j_0\right]\right) = g_0\frac{\frac{1}{g_0\xi^2}\frac{\partial}{\partial \xi}\left[g_0\xi^2\left(\sigma_{\rm n}^2\xi-\tilde{F}_{\rm n}\right)\right]-\frac{\ell\left(\ell+1\right)}{\xi^2}\left(\tilde{j}_{\rm n}-\frac{3\left(\gamma-4/3\right)}{\gamma-1}\frac{\gamma h_0}{g_0}\right)}{\sigma_{\rm n}^2+\frac{\ell\left(\ell+1\right)}{\xi^2}\frac{\gamma h_0}{g_0}}. \label{jnspherical}
\end{equation}
Here we normalized the unperturbed gravitational potential such that $j_0(1) = 0$, which, as for the cylindrical case, does not affect the solutions but does simplify the appearance of the equations. This set of coupled equations is the analog of Equations \eqref{Feigentot} and \eqref{jeigentot}, and we note that -- aside from geometrical factors, slight differences in the dependence on the adiabatic index and the unperturbed potential, and the appearance of $\ell(\ell+1)/\xi^2$ as opposed to $k^2+m^2/\xi^2$ -- the two sets of equations are identical. As we did in the cylindrical case, we can now expand the solutions near the surface of the polytrope to determine the boundary conditions on the functions at $\xi = 1$. Requiring that the normal component of the velocity be continuous in the comoving frame of the surface and expanding the gravitational potential as

\begin{equation}
\tilde{j}_{\rm n}(\xi \simeq 1) = J_{\rm n}+J_{\rm n}'\left(1-\xi\right),
\end{equation}
the series expansion of Equation \eqref{Fnspherical} near the surface gives

\begin{equation}
\frac{\partial \tilde{F}_{\rm n}}{\partial \xi}\bigg{|}_{\xi = 1} = -\frac{\sigma_{\rm n}^2}{\gamma}\left(\sigma_{\rm n}^2+2\left(\gamma-2\right)+\frac{1}{\sigma_{\rm n}^2}\ell\left(\ell+1\right)\left(\gamma-1\right)J_{\rm n}-J_{\rm n}'\right).
\end{equation}
Outside of the surface of the polytrope, Equation \eqref{jnspherical} is Laplace's equation in spherical coordinates for the quantity $\tilde{j}_{\rm n}-\partial/\partial\xi\left[\xi j_0\right]$. Requiring that the perturbation to the gravitational potential remain finite at large radii then yields the following, additional constraint between $J_{\rm n}'$ and $J_{\rm n}$:

\begin{equation}
J_{\rm n}' = \left(\ell+1\right)\left(J_{\rm n}-1\right).
\end{equation}
We can now numerically solve this set of equations in a manner analogous to what was done for the cylindrical eigenmodes, and search for the pairs of quantities $\{\sigma_{\rm n}, J_{\rm n}\}$ that simultaneously satisfy the regularity of both the velocity and the gravitational potential near the origin; these are the eigenmodes that describe the fundamental oscillations of the spherical polytrope. Using this method, we find precisely the same eigenvalues (to four significant figures) as reported in \citet{lee86} for the eigenmodes of a $\gamma = 5/3$ polytrope. Namely, the $f$-mode has $\sigma_{\rm n}^2 \simeq -2.12$ and $J_{\rm n} = 0.796$, and the higher-order modes are the same as those reported in \citet{lee86} with $J_{\rm n} \simeq 1$ (as for the cylindrical oscillations, the feature $J_{\rm n} \simeq 1$ is a direct demonstration of the validity of Cowling's approximation).

\bibliographystyle{aasjournal}
\bibliography{refs}

\end{document}